\documentclass[ALICE,manyauthors]{cernphprep}
\usepackage[comma,square,numbers,sort&compress]{natbib}

\usepackage{lineno}
\usepackage{xspace}
\usepackage{hyperref}
\usepackage[dvipsnames]{xcolor}
\usepackage{orcidlink}
\usepackage[T1]{fontenc}

\begin{document}
%%%%%%%%%%%%%%%%%%%%%%%%%%%%%%%%%%%%%%%%%%%%%%%%%%
% These are some new commands that may be useful 
% for paper writing in general. If other newcommands
% are needed for your specific paper, please feel 
% free to add here. 
%
% The currently available commands are organized in: 
% 1) Systems
% 2) Quantities
% 3) Energies and units
% 4) Detectors
% 5) particle species 
%%%%%%%%%%%%%%%%%%%%%%%%%%%%%%%%%%%%%%%%%%%%%%%%%%

% 1) SYSTEMS 
\newcommand{\pp}           {pp\xspace}
\newcommand{\ppbar}        {\mbox{$\mathrm {p\overline{p}}$}\xspace}
\newcommand{\XeXe}         {\mbox{Xe--Xe}\xspace}
\newcommand{\PbPb}         {\mbox{Pb--Pb}\xspace}
\newcommand{\pA}           {\mbox{pA}\xspace}
\newcommand{\pPb}          {\mbox{p--Pb}\xspace}
\newcommand{\AuAu}         {\mbox{Au--Au}\xspace}
\newcommand{\dAu}          {\mbox{d--Au}\xspace}

% 2) QUANTITIES 
\newcommand{\s}            {\ensuremath{\sqrt{s}}\xspace}
\newcommand{\snn}          {\ensuremath{\sqrt{s_{\mathrm{NN}}}}\xspace}
\newcommand{\pt}           {\ensuremath{p_{\rm T}}\xspace}
\newcommand{\meanpt}       {$\langle p_{\mathrm{T}}\rangle$\xspace}
\newcommand{\ycms}         {\ensuremath{y_{\rm CMS}}\xspace}
\newcommand{\ylab}         {\ensuremath{y_{\rm lab}}\xspace}
\newcommand{\etarange}[1]  {\mbox{$\left | \eta \right |~<~#1$}}
\newcommand{\yrange}[1]    {\mbox{$\left | y \right |~<~#1$}}
\newcommand{\dndy}         {\ensuremath{\mathrm{d}N_\mathrm{ch}/\mathrm{d}y}\xspace}
\newcommand{\dndeta}       {\ensuremath{\mathrm{d}N_\mathrm{ch}/\mathrm{d}\eta}\xspace}
\newcommand{\avdndeta}     {\ensuremath{\langle\dndeta\rangle}\xspace}
\newcommand{\dNdy}         {\ensuremath{\mathrm{d}N_\mathrm{ch}/\mathrm{d}y}\xspace}
\newcommand{\Npart}        {\ensuremath{N_\mathrm{part}}\xspace}
\newcommand{\Ncoll}        {\ensuremath{N_\mathrm{coll}}\xspace}
\newcommand{\dEdx}         {\ensuremath{\textrm{d}E/\textrm{d}x}\xspace}
\newcommand{\RpPb}         {\ensuremath{R_{\rm pPb}}\xspace}

% 3) ENERGIES, UNITS
\newcommand{\nineH}        {$\sqrt{s}~=~0.9$~Te\kern-.1emV\xspace}
\newcommand{\seven}        {$\sqrt{s}~=~7$~Te\kern-.1emV\xspace}
\newcommand{\twoH}         {$\sqrt{s}~=~0.2$~Te\kern-.1emV\xspace}
\newcommand{\twosevensix}  {$\sqrt{s}~=~2.76$~Te\kern-.1emV\xspace}
\newcommand{\five}         {$\sqrt{s}~=~5.02$~Te\kern-.1emV\xspace}
\newcommand{\twosevensixnn}{$\sqrt{s_{\mathrm{NN}}}~=~2.76$~Te\kern-.1emV\xspace}
\newcommand{\fivenn}       {$\sqrt{s_{\mathrm{NN}}}~=~5.02$~Te\kern-.1emV\xspace}
\newcommand{\eightnn}       {$\sqrt{s_{\mathrm{NN}}}~=~8.16$~Te\kern-.1emV\xspace}
\newcommand{\LT}           {L{\'e}vy-Tsallis\xspace}
%original:
%\newcommand{\GeVc}         {Ge\kern-.1emV/$c$\xspace}
%\newcommand{\MeVc}         {Me\kern-.1emV/$c$\xspace}
%\newcommand{\TeV}          {Te\kern-.1emV\xspace}
%\newcommand{\GeV}          {Ge\kern-.1emV\xspace}
%\newcommand{\MeV}          {Me\kern-.1emV\xspace}
%\newcommand{\GeVmass}      {Ge\kern-.2emV/$c^2$\xspace}
%\newcommand{\MeVmass}      {Me\kern-.2emV/$c^2$\xspace}

\newcommand{\GeVc}         {GeV/$c$\xspace}
\newcommand{\MeVc}         {MeV/$c$\xspace}
\newcommand{\TeV}          {TeV\xspace}
\newcommand{\GeV}          {GeV\xspace}
\newcommand{\MeV}          {MeV\xspace}
\newcommand{\GeVmass}      {GeV/$c^2$\xspace}
\newcommand{\MeVmass}      {MeV/$c^2$\xspace}

\newcommand{\lumi}         {\ensuremath{\mathcal{L}}\xspace}

% 4) DETECTORS 
\newcommand{\ITS}          {\rm{ITS}\xspace}
\newcommand{\TOF}          {\rm{TOF}\xspace}
\newcommand{\ZDC}          {\rm{ZDC}\xspace}
\newcommand{\ZDCs}         {\rm{ZDCs}\xspace}
\newcommand{\ZNA}          {\rm{ZNA}\xspace}
\newcommand{\ZNC}          {\rm{ZNC}\xspace}
\newcommand{\SPD}          {\rm{SPD}\xspace}
\newcommand{\SDD}          {\rm{SDD}\xspace}
\newcommand{\SSD}          {\rm{SSD}\xspace}
\newcommand{\TPC}          {\rm{TPC}\xspace}
\newcommand{\TRD}          {\rm{TRD}\xspace}
\newcommand{\VZERO}        {\rm{V0}\xspace}
\newcommand{\VZEROA}       {\rm{V0A}\xspace}
\newcommand{\VZEROC}       {\rm{V0C}\xspace}
\newcommand{\Vdecay} 	   {\ensuremath{V^{0}}\xspace}

% 4) PARTICLE SPECIES 
\newcommand{\jpsi}         {J/$\psi$\xspace}
\newcommand{\ee}           {\ensuremath{\rm{e}^{+}\rm{e}^{-}}\xspace} 
\newcommand{\pip}          {\ensuremath{\pi^{+}}\xspace}
\newcommand{\pim}          {\ensuremath{\pi^{-}}\xspace}
\newcommand{\kap}          {\ensuremath{\rm{K}^{+}}\xspace}
\newcommand{\kam}          {\ensuremath{\rm{K}^{-}}\xspace}
\newcommand{\pbar}         {\ensuremath{\rm\overline{p}}\xspace}
\newcommand{\kzero}        {\ensuremath{{\rm K}^{0}_{\rm{S}}}\xspace}
\newcommand{\lmb}          {\ensuremath{\Lambda}\xspace}
\newcommand{\almb}         {\ensuremath{\overline{\Lambda}}\xspace}
\newcommand{\Om}           {\ensuremath{\Omega^-}\xspace}
\newcommand{\Mo}           {\ensuremath{\overline{\Omega}^+}\xspace}
\newcommand{\X}            {\ensuremath{\Xi^-}\xspace}
\newcommand{\Ix}           {\ensuremath{\overline{\Xi}^+}\xspace}
\newcommand{\Xis}          {\ensuremath{\Xi^{\pm}}\xspace}
\newcommand{\Oms}          {\ensuremath{\Omega^{\pm}}\xspace}
\newcommand{\degree}       {\ensuremath{^{\rm o}}\xspace}

% 5) CONVENIENCE
\newcommand{\ccBar}          {\ensuremath{\rm c\overline{c}}\xspace}
\newcommand{\bbBar}          {\ensuremath{\rm b\overline{b}}\xspace}
\newcommand{\MeanPt}         {\ensuremath{\langle\pt\rangle}\xspace}
\newcommand{\MeanPtSquare}         {\ensuremath{\langle\pt^{2}\rangle}\xspace}
\newcommand{\der}            {\ensuremath{\text{d}}}
\newcommand{\fb} {$f_{\rm b}$\xspace}       

%%%%%%%%%%%%%%%  Title page %%%%%%%%%%%%%%%%%%%%%%%%
\begin{titlepage}
% the dates below correspond to CERN approval
% please don't touch: EB chairs will take care
\PHyear{2022}       % required, will be obtained from CERN
\PHnumber{256}      % required, will be obtained from CERN
\PHdate{17 November}  % required, will be obtained from CERN
%%%%%%%%%%%%%%%%%%%%%%%%%%%%%%%%%%%%%%%%%%%%%%%%%%%%

%%% Put your own title + short title here:
\title{J/$\psi$ production at midrapidity in p$-$Pb collisions at $\mathbf{\sqrt{{\textsl{s}}_{ NN}} = 8.16}$ TeV}
\ShortTitle{J/$\psi$ production at midrapidity in p$-$Pb collisions at $\sqrt{s_{\rm NN}} = 8.16$ TeV}   % appears on left page headers
 
%%% Do not change the next lines
\Collaboration{ALICE Collaboration\thanks{See Appendix~\ref{app:collab} for the list of collaboration members}}
\ShortAuthor{ALICE Collaboration} % appears on right page headers, do not change

\begin{abstract}

The production of inclusive, prompt and non-prompt \jpsi was studied for the first time at midrapidity ($ -1.37 < y_{\rm cms} < 0.43$) in \pPb collisions at $\sqrt{s_{\rm NN}} = 8.16$~TeV with the ALICE detector at the LHC. The inclusive \jpsi mesons were reconstructed in the dielectron decay channel in the transverse momentum (\pt) interval $0 < \pt < 14$~\GeVc and the prompt and non-prompt contributions were separated on a statistical basis for $\pt > 2$~\GeVc. The study of the \jpsi mesons in the dielectron channel used for the first time in ALICE online single-electron triggers from the Transition Radiation Detector, providing a data sample corresponding to an integrated luminosity of $689 \pm 13~ \mu{\rm b}^{-1}$. The proton$-$proton reference cross section for inclusive \jpsi was obtained based on interpolations of measured data at different centre-of-mass energies and a universal function describing the \pt-differential \jpsi production cross sections. The \pt-differential nuclear modification factors \RpPb of inclusive, prompt, and non-prompt \jpsi are consistent with unity and described by theoretical models implementing only nuclear shadowing.

\end{abstract}
\end{titlepage}

\setcounter{page}{2} %please do not remove this line

%%%%%%%%%%%%%%%%%%%%%%%%%%%%%%%%
% begin main text
%%%%%%%%%%%%%%%%%%%%%%%%%%%%%%%%

\section{Introduction} 

Differential measurements of \jpsi mesons in heavy-ion collisions at the LHC give insight into their production mechanisms, pointing to a significant contribution from regeneration at low \pt, and give evidence for deconfinement in Pb--Pb collisions~\cite{ALICE:2019nrq,ALICE:2019lga,ALICE:2020pvw,Zhou:2014kka,Du:2015wha,Andronic:2019wva}. However, to better understand the underlying mechanisms and the influence of the quark$-$gluon plasma, reference measurements in proton$-$proton (pp) and proton$-$nucleus (p$-$A) collisions are crucial. Measurements in p$-$A collisions allow cold nuclear matter effects to be quantified. In \pPb collisions at the LHC, parton shadowing or gluon saturation are considered to be the dominant effects influencing \jpsi production. These initial-state effects are expressed in terms of modified parton distribution functions in the nucleus (nPDF) or the color glass condensate effective theory~\cite{Blaizot:2004wv,Armesto:2006ph,Gelis:2012ri}. Additional processes in the initial state, such as coherent parton energy loss, and final-state effects, where the $\ccBar$ states interact with the system generated in the small collision volume, have been predicted as well~\cite{Arleo:2013zua,Ferreiro:2014bia,Chen:2016dke,Du:2018wsj}. 

The inclusive \jpsi cross section includes contributions from prompt \jpsi, directly produced in the hadronic interaction or via feed-down from other directly produced charmonium states
(e.g.\ $\chi_{\rm c}$ and $\psi(2S)$), as well as non-prompt \jpsi originating from the decay of beauty hadrons. Thus, the influence of cold nuclear matter effects on open-heavy flavour production can be accessed as well through non-prompt \jpsi measurements (b-hadron  $\rightarrow$ \jpsi + X ) in \pPb collisions. 

The ALICE, ATLAS, CMS and LHCb collaborations performed many differential measurements of \jpsi and open-beauty production at mid-, forward and backward rapidity in \pPb collisions at \linebreak \fivenn~\cite{LHCb:2013gmv,ALICE:2015sru,ALICE:2015kgk,ALICE:2016uid,CMS:2017exb,ATLAS:2017prf,ALICE:2021lmn, CMS:2015sfx} as well as at forward and backward rapidity at \eightnn~\cite{LHCb:2017ygo,ALICE:2018mml,LHCb:2019avm}. All the measurements can qualitatively be described by theoretical calculations including different combinations of the above-mentioned effects.

This article reports, for the first time, on the measurement of inclusive, prompt and non-prompt \jpsi mesons at midrapidity in \pPb collisions at \eightnn. At low \pt, the measurements reach Bjorken-$x$ values of $10^{-4}$ to $10^{-3}$. As the ALICE trigger strategy for \pPb collisions at this energy only allocated a small portion of the bandwidth for minimum bias triggers, in order to maximise the live time for rare triggers, the presented studies were only possible thanks to the usage of single-electron triggers provided by the ALICE Transition Radiation Detector (TRD).

This article is organised as follows. A brief description of the ALICE detector with a focus on the detectors used for the analysis is given in Sec.~\ref{Section::Detector}, where the data sample and event selection are also discussed. The analysis details and the estimation of the systematic uncertainties are described in Sec.~\ref{Section::Analysis}. To quantify possible modifications of the \jpsi production in \pPb collisions, a reference \pt-differential cross section for pp collisions is obtained by interpolating measured data from different collision energies, as detailed in Sec.~\ref{Section::ppRef}. The results are presented and discussed in Sec.~\ref{Section::Results}. Finally, conclusions are drawn in Sec.~\ref{Section::Summary}. 

\section{Detector setup, data sample, and event selection}\label{Section::Detector}

The ALICE detector~\cite{ALICE:2008ngc,ALICE:2014sbx} is ideally suited to measure \jpsi production in the dielectron decay channel at midrapidity due to its low material budget as well as its excellent particle identification (PID) and transverse momentum resolution.  \\
The global track reconstruction is performed using the Inner Tracking System (ITS)~\cite{ALICE:2010tia} as well as the Time Projection Chamber (TPC)~\cite{Alme:2010ke}, both covering the full azimuthal angle and pseudorapidity $| \eta | < 0.9$. The detectors are placed inside a solenoid magnet, providing a magnetic field of $B = 0.5$~T along the beam direction. The ITS includes 6~layers of silicon detectors (2 layers each of silicon pixel, drift, and strip detectors) and is used for the primary and secondary vertex reconstruction. The silicon pixel layers (SPD), which are the two innermost ones, are placed at radial distances of $r = 3.9$ and 7.6~cm from the nominal interaction point (IP). Tracks with hits in both SPD layers (only in the second layer) have an impact parameter resolution better than 50~$\mu$m (100~$\mu$m) in the transverse plane for transverse momenta above 2~\GeVc, allowing the prompt and non-prompt \jpsi contributions to be separated on a statistical basis. The electron identification is based on the measurement of the specific ionisation energy loss (d$E$/d$x$) in the TPC, a cylindrical gaseous drift detector with dimensions 85~$< r < $~250~cm along the radial and $|z| <$~250~cm along the beam direction from the IP. 

To enrich samples of electrons and positrons at intermediate and high \pt, two online single-electron triggers\footnote{The TRD trigger selects online electrons and positrons. Throughout the article the term ‘electron trigger’ denotes both electron and positron.} derived from information provided by the TRD~\cite{ALICE:2017ymw} were deployed. The TRD consists of 522~chambers\footnote{Eighteen chambers are not installed in front of the PHOton Spectrometer~\cite{ALICE:1999ozr} to reduce the material budget in front of this detector.} arranged in 6~layers surrounding the TPC in full azimuth at a radial distance of 2.90~m to 3.68~m from the IP, and along the longitudinal direction in 5 stacks covering the pseudorapidity interval $  |\eta| < 0.84$. Each chamber comprises a foam/fibre sandwich radiator, a drift volume and a multiwire proportional chamber filled with a Xe-CO$\rm{ _{2}}$ gas mixture. 
The measurement of the temporal evolution of the signal in the drift region allows track segments to be reconstructed in each chamber, as well as the specific ionisation energy loss of the charged particle and the transition radiation photons from electrons crossing the radiator with a Lorentz factor $\gamma > 800$ to be measured. Due to the fast readout and subsequent online reconstruction of the TRD signals, where a transverse momentum and an electron likelihood\footnote{The total accumulated charge of each track segment is translated into an electron likelihood via a transformation function, stored in the form of a one-dimensional look-up table in the front-end electronics of the detector. The electron likelihood of a track is then obtained as the average of the likelihood values of the associated track segments.} is calculated within a stack from the individual track segments, a trigger decision on individual tracks with \pt and likelihood thresholds is made about 6~$\mu \rm{s}$ after the collision (level-1 trigger decision). 

Furthermore, two scintillator arrays (V0)~\cite{ALICE:2013axi} placed along the beam direction at $-3.7 < \eta < -1.7$ and $2.8 < \eta < 5.1$ are used for triggering and event characterisation. 

The analysis discussed in this article is based on data recorded in 2016 during the LHC heavy-ion run, where lead ions and protons were collided at a centre-of-mass energy per nucleon pair of \linebreak \eightnn. Data were taken with two beam configurations, where the directions of the proton and lead beams were swapped. With respect to the laboratory frame, the proton$-$lead centre-of-mass frame is shifted by $\Delta y_{\rm NN} = 0.465$ towards the incoming proton beam for both beam configurations, leading to a rapidity coverage of $ -1.37 < y_{\rm cms} < 0.43$. Minimum bias (MB) events (level-0) were selected based on the coincident signal of both V0 scintillator arrays. The trigger is fully efficient for recording events with inclusive \jpsi mesons. Only a small sample of MB events was recorded and used for the \jpsi analysis at low \pt and to evaluate the TRD trigger performance. The two single-electron triggers were run with \pt thresholds of 2 and 3 \GeVc and two different thresholds of the electron (positron) likelihood. The different efficiencies for electrons and positrons arise from the $E \times B$ effect\footnote{
In the chambers, the direction of the drift electric field is perpendicular to that of the magnetic field, which affects the direction of the drifting electrons such that it is approximately aligned with the negative tracks and systematically rotated relative to the positive tracks. 
This leads to differences in performance of the track segment reconstruction and the electron likelihood calculation.}~\cite{ALICE:2017ymw}. To reduce the background of electrons from photon conversions in the detector material, especially at large radii outside the TPC, 5~track segments per stack and a track segment in the first layer were required. In addition, a cut on the sagitta ($\Delta p_{\rm T}^{-1} < 0.2~c\rm{/GeV}$) was applied. These trigger selection criteria result in a reduction of the geometrical acceptance, because, due to individual inactive chambers, about 20\% of the stacks do not contribute to the trigger decision~\cite{ALICE:2017ymw}.

\begin{figure}[tb]
	\centering
  \includegraphics[width=.485\textwidth]{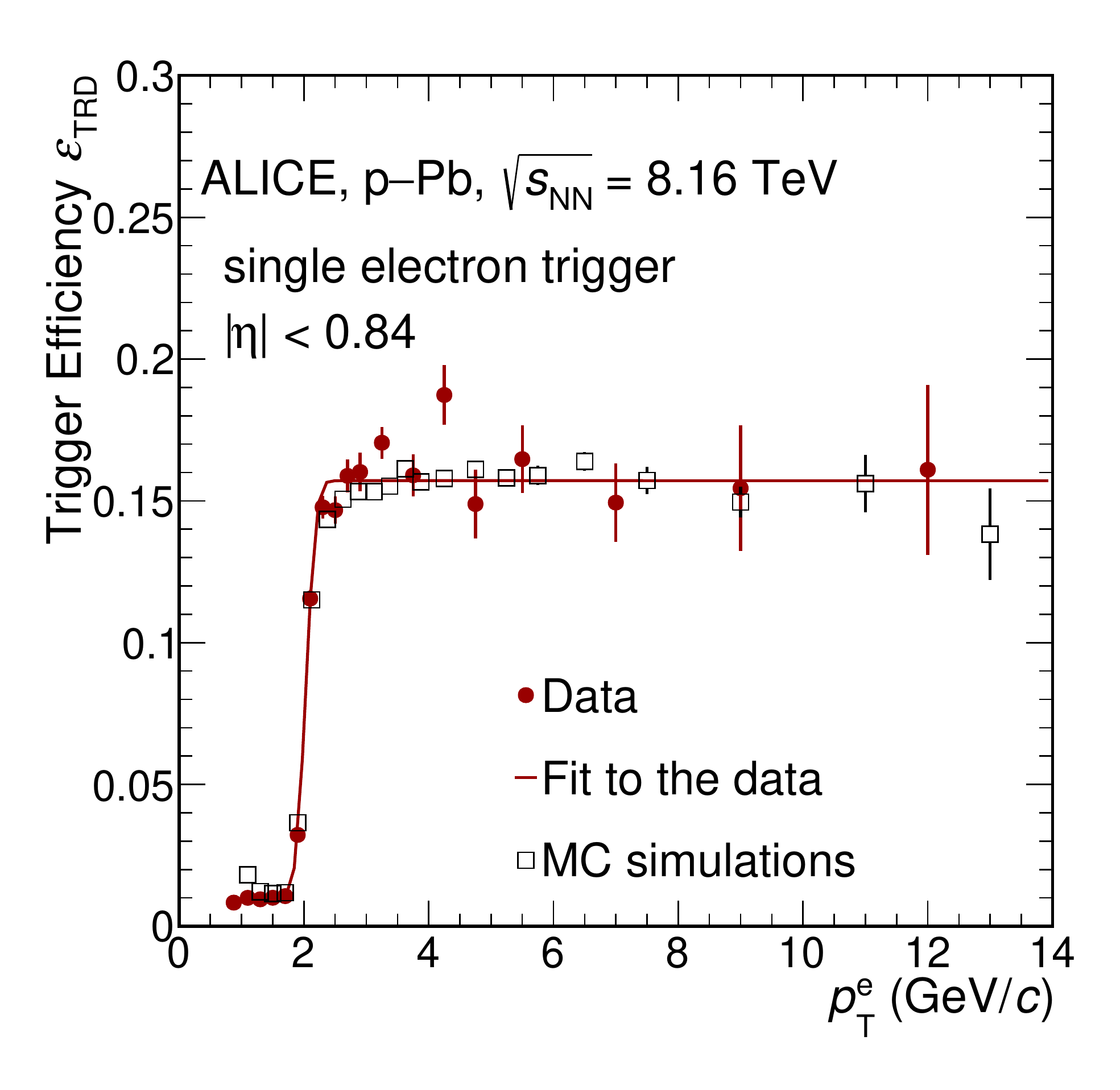}
  \includegraphics[width=.485\textwidth]{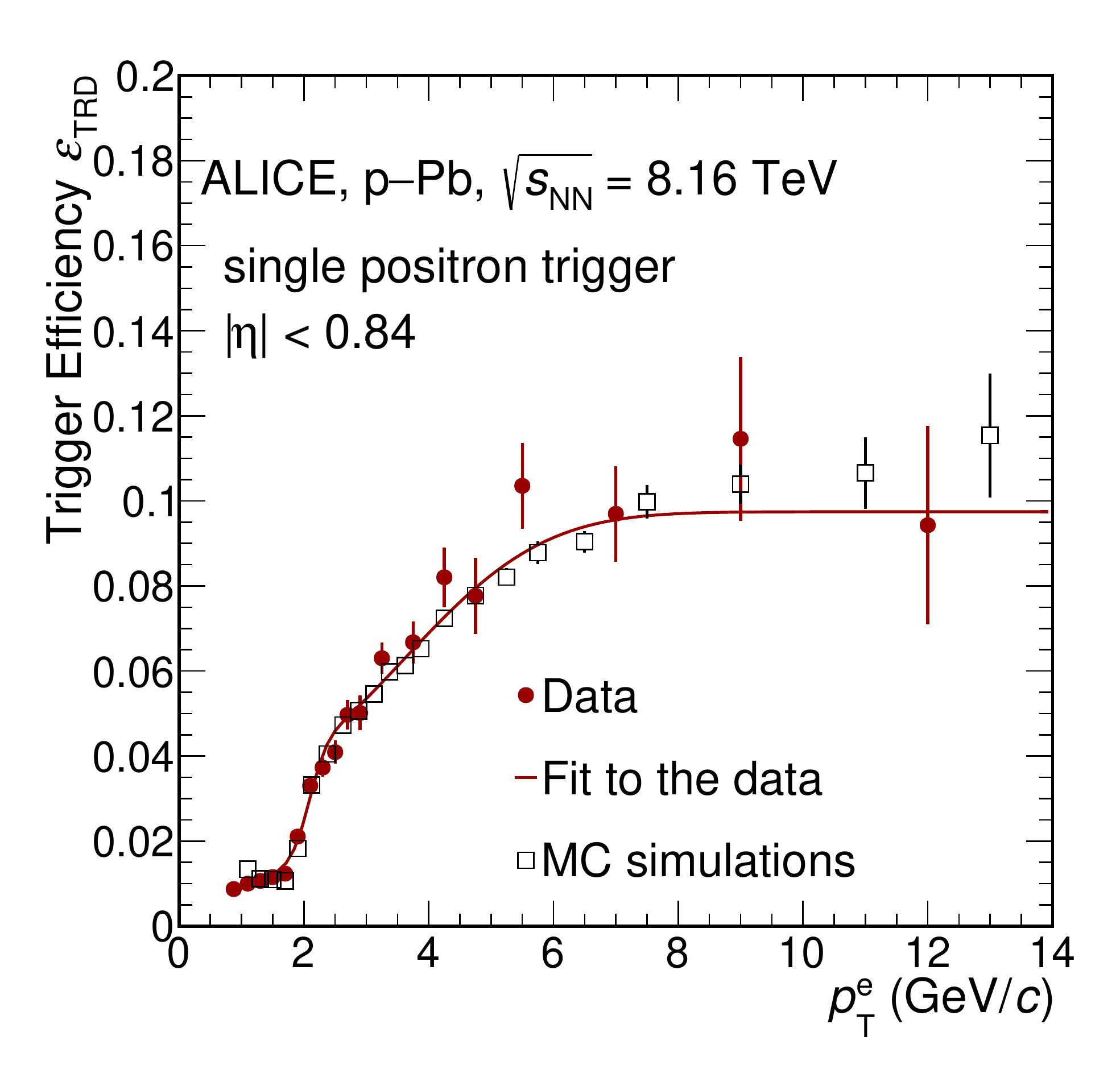}
	\caption{Acceptance times efficiency of the TRD trigger with the \pt threshold at 2~\GeVc for single electrons (left) and single positrons (right) obtained from data and MC simulations. See text for details.}
	\label{Fig1}
\end{figure}

Figure~\ref{Fig1} shows, as an example the acceptance times efficiency $\left (\epsilon_{\rm TRD} \right )$ of the TRD trigger with the online \pt threshold at 2~\GeVc for electrons (left) and positrons (right) considering all trigger selection criteria. The trigger turn-on curve results were obtained in the offline analysis as a function of the globally reconstructed \pt by identifying electrons and positrons using the TPC in MB data $\left (N_{\rm e}^{\rm MB}\right )$ and the fraction of these that satisfy the online TRD trigger decisions $\left (N_{\rm e}^{\rm MB,TRD}\right )$: $\epsilon_ {\rm TRD} (p_{\rm T}) = \frac{N_{\rm e}^{\rm MB,TRD}(p_{\rm T})} {N_{\rm e}^{\rm MB}(p_{\rm T})}$. The TRD trigger efficiency shows small variations as a function of time, as the gain of the gas detector depends on pressure and gas composition affecting both the electron and positron efficiencies. In addition, the downscaling factor of the MB trigger was modified during the data taking period. Thus, to compute the correct efficiencies for the measurement, the single-electron trigger efficiencies determined in MB data were studied as a function of time and weighted by the number of TRD-triggered events in each time interval. The \pt threshold is clearly visible as a sharp rise at 2~\GeVc, followed by a constant plateau at an efficiency value determined by the geometrical and the electron identification selection criteria outlined above. The entries at low \pt are either electrons or positrons in events that were recorded because a high-\pt electron or positron in the same event satisfied the online trigger condition. The difference in the trigger turn-on curves for electrons and positrons is due to the $E \times B$ effect and it is well reproduced in Monte Carlo (MC) simulations (see next section). To guide the eye, the turn-on curve of electrons is fitted by an error function with an offset to describe the underlying event. For positrons, another error function is added to better describe the slower rise of the distribution. 

Events selected by either of the two single-electron triggers are considered for further analysis, and to ensure a uniform detector acceptance, only events with a primary vertex located within 10~cm from the IP along the beam direction are accepted. Beam$-$gas interactions and pile-up events were rejected offline using information from the V0 and SPD detectors, together with algorithms identifying more than one vertex within an event, as described in Ref.~\cite{ALICE:2014sbx}.
The remaining fraction of pile-up events is negligible. These selection criteria result in MB and TRD-triggered data samples of about 50 million and 10 million events, corresponding to an integrated luminosity of $24.2  \pm 0.5~ \mu{\rm b}^{-1}$ and $689 \pm 13~ \mu{\rm b}^{-1}$, respectively.

\section{Data analysis}\label{Section::Analysis}

The measurements of inclusive, prompt, and non-prompt \jpsi production were performed in a similar way to previous analyses in pp and \pPb collisions at other centre-of-mass energies~\cite{ALICE:2021dtt,ALICE:2019pid,ALICE:2021edd,ALICE:2021lmn}. However, in this article TRD-triggered events were used for the first time for the \jpsi analysis. The minimum \jpsi \pt of the TRD-triggered analysis is 2~\GeVc, where the single-electron efficiencies result in a \jpsi efficiency of about 10\% increasing with rising \pt, as shown in Sec.~\ref{inclusiveanalysis}. For the measurement in the \pt~interval 0$-$2~\GeVc, the small available MB data sample was used.  

In a first step, tracks with good quality were selected, which in addition had to fulfil the requirement of a minimum transverse momentum of 1~\GeVc and a pseudorapidity of $ |\eta| < 0.84$ to reject tracks outside the TRD kinematic acceptance. For the inclusive analysis, to reduce the background originating from photon conversions at larger radii, tracks were required to have a hit in the first SPD layer. Also, in order to further suppress this contamination in the inclusive analysis, electrons (positrons) paired with positrons (electrons) in the same event yielding an invariant mass below 0.05 \GeVmass were rejected. The requirements on the SPD layers for the separation of the prompt and non-prompt \jpsi are outlined in Sec.~\ref{Nonpromptanalysis}.

The electron and positron identification is based on the measurement of the specific ionisation energy loss in the TPC. The selection criterion is $n_{\sigma_{\rm i}}$, which is defined as the difference between the measured and expected signal in units of the detector resolution, for a specific particle hypothesis $(i)$. Particles satisfying the condition $| n_{\sigma_{\rm e}} |< 3.0$ were thus identified as electrons.
Pions and protons were rejected by excluding tracks that were compatible within $3~n_{\sigma_{\rm \pi, p}}$ with the corresponding particle hypothesis. For tracks with momenta $p > 5$~\GeVc, the rejection criterion was reduced to $2~n_{\sigma_{\rm \pi, p}}$ to increase the \jpsi reconstruction efficiency at high \pt. For the MB analysis, the selection criterion for electrons and positrons was set to $ -2 < n_{\sigma_{\rm e}} < 3$ to improve the signal-to-background ratio at low~\pt.  

The efficiencies of the applied selection criteria and of the TRD trigger were obtained from MC simulations. The EPOS-LHC model~\cite{Pierog:2013ria} was used to simulate minimum bias \pPb collisions, into which one \jpsi meson per event was embedded. The prompt \jpsi mesons were generated with a flat rapidity distribution and a realistic \pt distribution taken from the \jpsi measurement in the dimuon decay channel at forward rapidity at the same centre-of-mass energy~\cite{ALICE:2018mml}. For the non-prompt \jpsi, PYTHIA~6.4~\cite{Sjostrand:2006za} was used to generate the \bbBar pairs hadronising into beauty hadrons, subsequently forced to decay into \jpsi.
The \jpsi decays were then simulated using the EvtGen package~\cite{Lange:2001uf} together with the PHOTOS model~\cite{Barberio:1993qi} for a proper description of the QED radiative decay channel (\jpsi~$\rightarrow \rm{e^+e^-}\gamma$). The MC simulation assumes the prompt \jpsi to be unpolarised, while the non-prompt \jpsi have a small residual polarisation arising from the contributions of the different b-hadron decay channels as implemented in EvtGen~\cite{Lange:2001uf}.
All generated particles were transported through the ALICE detector setup using a GEANT3 model~\cite{Brun:1082634} considering a realistic detector response and reproducing the detector performance during the data taking. The TRD trigger was emulated in the simulation, i.e.\ the same selection criteria were calculated and applied as in real data.

\subsection{Inclusive \jpsi analysis}\label{inclusiveanalysis}

\begin{figure}[tb]
	\centering
  \includegraphics[clip,width=1\textwidth]{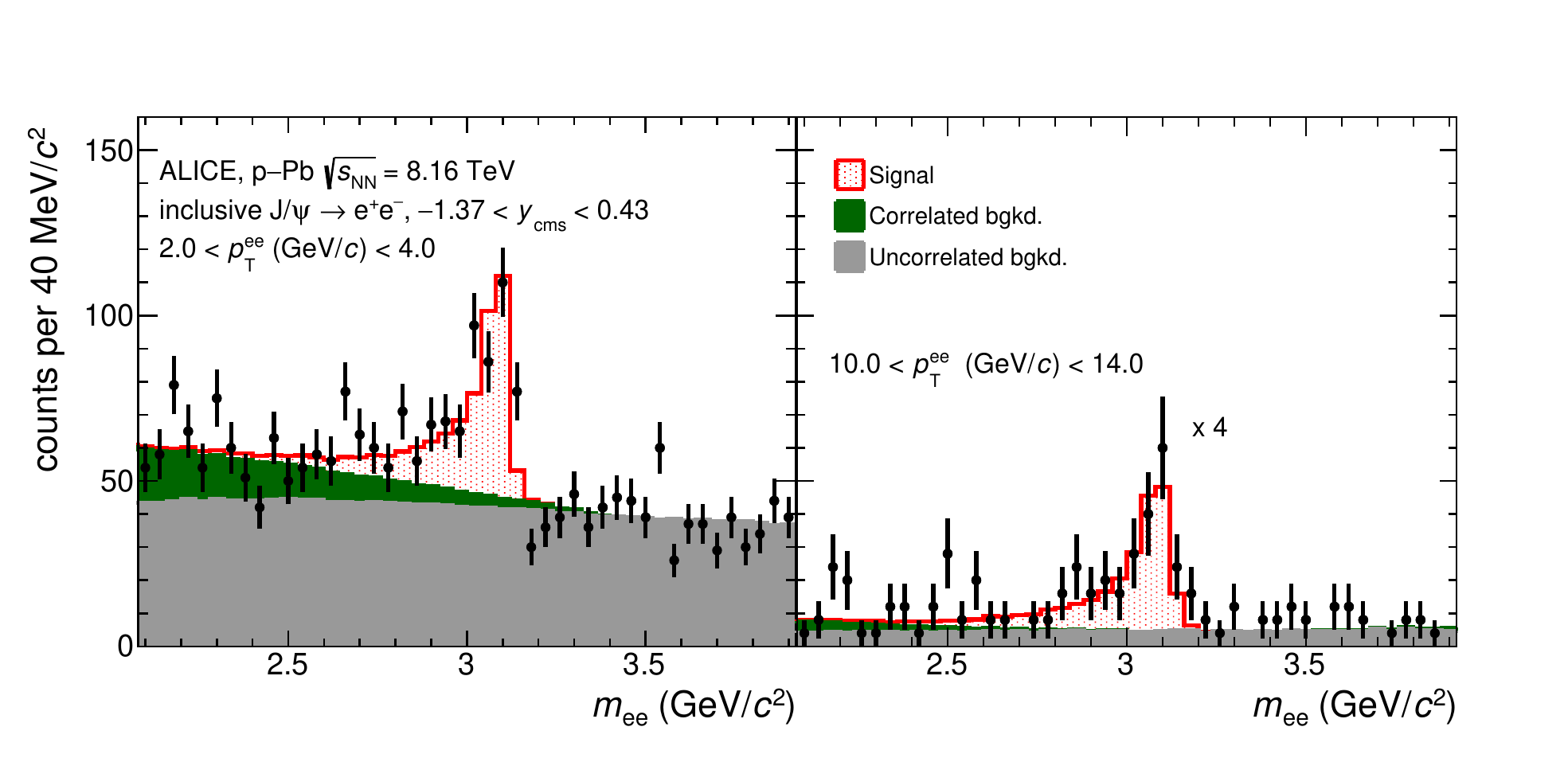}
	\caption{Invariant mass distribution of $\rm e^+ e^-$ pairs from \jpsi decays and from correlated and uncorrelated background sources for the lowest (left) and the highest (right) \pt interval considered in the analysis. For the highest \pt interval, the distributions were scaled by a factor~4 for better visibility.
	}
	\label{Fig2}
\end{figure}

The number of raw \jpsi candidates was extracted in \pt intervals from the invariant mass distribution ($m_{\rm ee}$) of electron$-$positron pairs after background subtraction. Figure~\ref{Fig2} shows the invariant mass distribution before background subtraction for two illustrative \pt intervals. The background is composed of pairs of electrons and positrons with different physics origin (uncorrelated background) and to a small extent, of electrons from common sources such as \ccBar and \bbBar decays or jet fragmentation (correlated background). Both background sources, see Fig.~\ref{Fig2}, are estimated by means of a hybrid method using the mixed-event technique (ME) for the uncorrelated background and a fitting function for the residual background, as applied in the analyses of minimum bias pp collisions at $\sqrt{s} = 5$ and 13~TeV~\cite{ALICE:2019pid,ALICE:2021dtt}.  The mixed-event background distribution was normalised in the mass interval $2 < m_{\rm ee} < 4$~\GeVmass to the measured distribution of the same-event like-sign pairs, as these are expected to be little affected by correlated sources. After subtraction of the ME background, the residual distribution was then fitted with a polynomial of second order and a MC template for the \jpsi signal. Counting the number of electron$-$positron pairs in the mass interval $2.92 < m_{\rm ee} < 3.16$~\GeVmass, after subtracting all background sources, yields the raw number of \jpsi candidates.

The \pt-differential cross section was obtained by correcting the number of \jpsi candidates found in a given transverse momentum ($\Delta \pt$) and rapidity ($\Delta y$) interval by the average acceptance and efficiency $\left( \langle Acc \times \epsilon_{\rm reco} \times \epsilon_{\rm mass}\times \epsilon_{\rm TRDtrg} \rangle \right)$ in these intervals:
\begin{equation}
     \frac{{\rm d}\sigma^{2}}{{\rm d}y{\rm d}p_{\rm T}} = \frac{N_{J/\psi}^{\rm raw}}{\left(\langle Acc \times \epsilon_{\rm reco} \times \epsilon_{\rm mass}\times \epsilon_{\rm TRDtrg} \rangle \right) \times {\rm BR} \times \Delta y \Delta p_{\rm T} \times \mathcal{L}_{\rm int}},
\end{equation}
where BR denotes the branching ratio of \jpsi to dielectrons (5.97 $\pm$ 0.03)\%~\cite{Zyla:2020zbs}. The integrated luminosity of the data sample is given by $\mathcal{L}_{\rm int} = \frac{N_{\rm MB}}{\sigma_{\rm MB}} $, where $\sigma_{\rm MB}$ is the MB trigger cross section obtained from van der Meer scans~\cite{ALICE-PUBLIC-2018-002}. The number of MB events $N_{\rm MB} = N_{\rm TRD} \times f_{\rm norm} $ was calculated from the number of TRD-triggered events $ N_{\rm TRD}$ and the normalisation factor $f_{\rm norm}$. The latter corresponds to the inverse probability of an MB event being triggered by the TRD as well. Its statistical uncertainty amounts to 0.35\%. The evaluation of $f_{\rm norm}$ was cross-checked with an alternative method based on the online counters, provided by the central trigger processor, of the number of inspected events at level-0 (MB) and level-1 (TRD-triggered). The ratio of the two numbers is corrected for the slightly different efficiency in event selection in both data samples. Due to the larger number of level-0 trigger counts, this ratio has a smaller statistical uncertainty and agrees on the few per-mill level with the result from the first method. Thus the statistical uncertainty of the first described method is used as the systematic uncertainty of the normalisation factor $f_{\rm norm}$. For the MB analysis, neither the TRD trigger efficiency nor the trigger normalisation factor $f_{\rm norm}$ apply.

\begin{figure}[tb]
	\centering
 
   \includegraphics[width=0.475\textwidth]{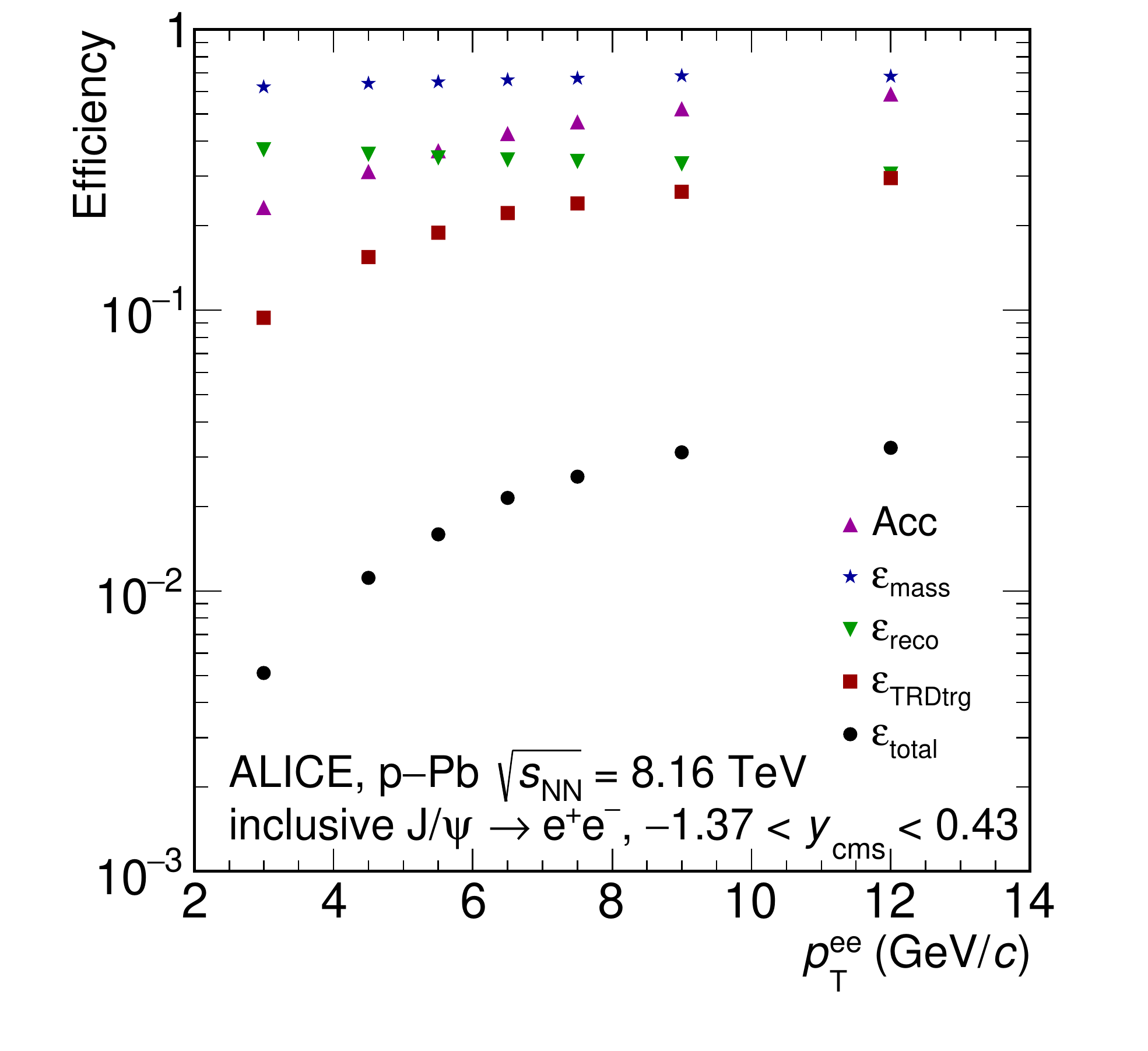}
  \hspace{-0.5cm}
     \includegraphics[width=0.49\textwidth]{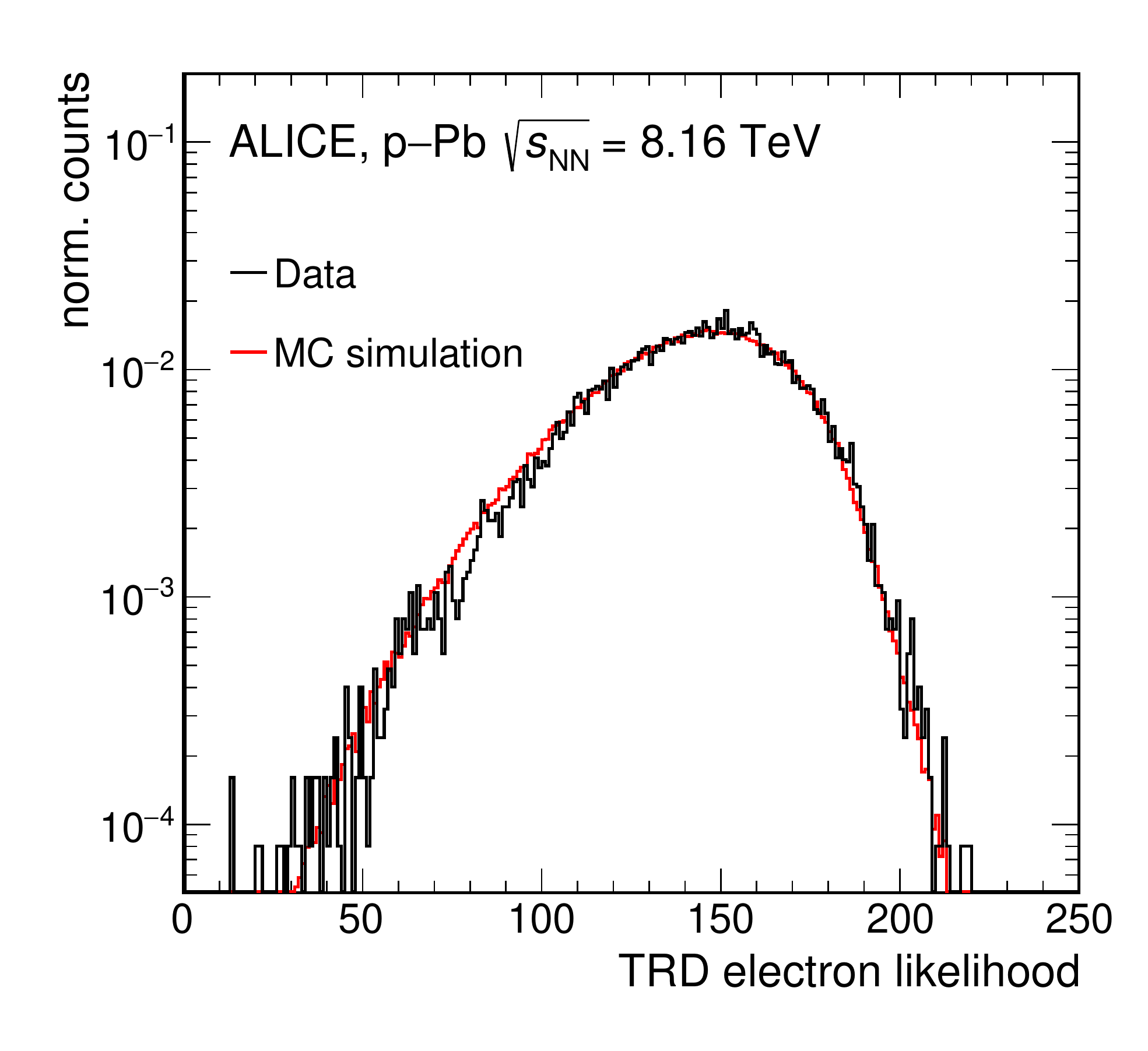}
	\caption{(Left) Efficiency as a function of transverse momentum for the inclusive \jpsi analysis of the TRD-triggered data. (Right) Electron likelihood estimated from TRD for electrons identified using the TPC particle identification capabilities in data and MC simulations. The electron likelihood is stored in hardware as an unsigned 8-bit value (translating into integer values from 0 to 255).}
	\label{Fig3}
\end{figure}

The acceptance and efficiency factors, determined via the aforementioned MC simulations, correct for the kinematic acceptance ($Acc$), the reconstruction efficiency and the applied selection criteria ($\epsilon_{\rm reco}$), the TRD trigger efficiency for \jpsi mesons ($\epsilon_{\rm TRDtrg}$) and the mass interval chosen to count the \jpsi signal candidates ($\epsilon_{\rm mass}$). The individual contributions and the total efficiency are shown in Fig.~\ref{Fig3} (left). The TRD trigger algorithm was emulated in the simulation. The same quantities such as electron likelihood, online \pt, and number of track segments per stack as in real data were calculated and the analogous trigger decision was derived. By comparing the distributions of these quantities for electrons and positrons in data and MC simulations, it was ensured that the TRD trigger is correctly implemented in simulation and shows the same performance as in data. Figure~\ref{Fig3} (right) shows as an example the excellent agreement of the electron and positron likelihood in data and MC simulations for the single-electron trigger with a \pt~threshold at 2~\GeVc. 

The following sources of systematic uncertainties were considered for the determination of the inclusive \jpsi cross section: (i) track reconstruction efficiency, (ii) electron identification, (iii) signal extraction, (iv) kinematics of the \jpsi used in the MC simulation, and (v) TRD trigger efficiency (does not apply to the analysis of the MB data sample). 
The uncertainty of the track reconstruction efficiency is related to the ITS$-$TPC matching efficiency and to the track selection criteria. No statistically significant systematic effects were found when rerunning the analysis with variations of the track selection criteria, with the exception of the requirement of the number of hits in the SPD layers (hit in first or both layers of the SPD). To estimate the influence of the SPD criterion, a data-driven technique, where pions tagged as belonging to identified \kzero decay topologies were used to determine the single-track uncertainty. The latter was then propagated to the two-track level (\jpsi) using a phase space simulation of the \jpsi decay to dielectrons. This uncertainty was found to amount to 3.1\%, independent of the pair \pt. It was verified using MC simulations that the single-track uncertainty obtained with pions is identical to the one of electrons. Likewise, the uncertainty of the ITS$-$TPC matching efficiency, describing the probability that a track reconstructed in the TPC also has matching hits in the ITS, was estimated in a data-driven procedure as described in~\cite{ALICE:2017olh} and found to be independent of particle species. The uncertainty was propagated to the two track level (\jpsi) and amounts to 2\%, independent of \pt. The two uncertainties were added in quadrature and are considered to be correlated across \pt intervals. \\
The uncertainty due to the particle identification was determined using electrons from photon conversions, pions from \kzero and protons from $\Lambda$ decays, topologically reconstructed in data and MC simulations. The comparisons of the electron identification efficiency and hadron rejection in data and MC simulations under variations of the PID selection criteria yield a 2\% \pt-independent uncertainty for the \jpsi meson. The uncertainty is considered correlated across \pt intervals. \\
The raw \jpsi yield was corrected for the average acceptance and efficiency in a given \pt interval and is thus sensitive to the kinematic distribution of the inclusive \jpsi mesons used in the MC simulation. To estimate the related uncertainty, the \pt spectrum of \jpsi mesons in the same rapidity range in \pPb at \fivenn~\cite{ALICE:2021lmn} was fitted with a power law; then, new distributions were derived by varying the obtained parameters according to the correlation matrix provided by the fit procedure. For each iteration the average acceptance and efficiency in the given \pt interval was recomputed and the RMS of all values with respect to the default value was taken as uncertainty. In most of the \pt intervals the uncertainty is negligible; the largest value, 0.2\%, is found in the lowest \pt interval.  \\
The signal extraction uncertainty is composed of contributions from the background description and the \jpsi signal distribution. The uncertainty related to the background description was determined by choosing different fitting functions for the correlated and uncorrelated background. The studies were performed in two \pt intervals (2$-$6 and 6$-$14~\GeVc) with similar background distributions to avoid the statistical uncertainties dominating the systematic uncertainties. The variations show a deviation with respect to the default value of 1.9\% for both \pt intervals. The uncertainty associated with the \jpsi signal shape was found by varying the mass interval used to count the raw \jpsi candidates. To reduce statistical fluctuations, the root mean square of all values with respect to the central value was determined for the \pt-integrated case resulting in an uncertainty of 1.3\%.  \\
The uncertainty of the TRD trigger efficiency was estimated based on MC simulations, where the threshold of the electron and positron likelihood value for the trigger decision was varied by an amount corresponding to the granularity with which it is calculated in the front-end electronics, and then the trigger efficiency for \jpsi mesons was recalculated. The resulting uncertainty is 2.3\%. The adopted threshold variation is larger than the one expected from the observed changes in pressure and gas composition, that would lead to a change in gain and thus a change in the electron efficiency of the trigger. \\
The systematic uncertainties of all sources described above were added in quadrature and amount to a total systematic uncertainty of 5.2\% for all \pt intervals. The total systematic uncertainty of the MB analysis is 4.7\%. The branching ratio uncertainty, the MB trigger cross section uncertainty (including also a contribution from its stability over time, as discussed in Ref.~\cite{ALICE-PUBLIC-2018-002}), and the uncertainty of the TRD trigger normalisation factor $f_{\rm norm}$ are added in quadrature to obtain a global normalisation uncertainty of 2\%. As the latter contribution is small, this uncertainty also holds for the MB analysis.

\subsection{Determination of the non-prompt \jpsi fraction}\label{Nonpromptanalysis}

The non-prompt \jpsi fraction (\fb) was determined as in previous analyses~\cite{ALICE:2012vpz,ALICE:2015nvt,ALICE:2018szk,ALICE:2021edd,ALICE:2021lmn} on a statistical basis. The method relies on the property that \jpsi mesons originating from b-hadron decays have, in the studied kinematic range, a decay vertex distribution extending to values well beyond the secondary vertex resolution, in contrast to prompt \jpsi. As a very good pointing resolution is needed, only \jpsi candidates with at least one of the decay products having hits in both SPD layers were accepted.

The measurement of the fraction \fb was carried out via a minimisation of a two-dimensional unbinned negative log-likelihood fit in \pt intervals, where the invariant mass and the pseudoproper decay length ($x$) distributions of the electron-positron pairs were simultaneously fitted:
\begin{equation}
    - {\rm ln}~L~=~- \sum_{i=1}^{N} {\rm ln}~F(x,m_{\rm ee}).
\end{equation}
 The variable $N$ denotes the number of \jpsi candidates in the invariant mass interval $2.4 < m_{\rm ee} < 3.6$~\GeVmass and $F(x,m_{\rm ee})$ is given as
\begin{equation}
     F(x,m_{\rm ee})~=~ f_{\rm sig} \times F_{\rm sig}(x) \times M_{\rm sig}(m_{\rm ee})~+~ (1-f_{\rm sig}) \times F_{\rm bkgd}(x) \times M_{\rm bkgd}(m_{\rm ee}),
\end{equation}
where $f_{\rm sig}$ is the fraction of \ee pairs attributed to prompt and non-prompt \jpsi within the invariant mass interval $2.4 < m_{\rm ee} < 3.6$~\GeVmass. $M_{\rm sig}(m_{\rm ee})$ and $F_{\rm sig}(x)$ are the functional forms describing the invariant mass and pseudoproper decay length distributions of the signal. $M_{\rm bkgd}(m_{\rm ee})$ and $F_{\rm bkgd}(x)$ are the corresponding functional forms of the background component. The pseudoproper decay length is defined as 
$x~=~\frac{c \times L_{\rm{xy}} \times m_{{\rm J}/\psi} }{\pt}$, where $ L_{\rm{xy}}$ is the transverse projection of the vector from the primary vertex of the event to the \jpsi decay vertex and $m_{{\rm J}/\psi}$ is the \jpsi pole mass~\cite{Zyla:2020zbs}. 
The signal is composed of a prompt and non-prompt \jpsi contribution

\begin{equation}
    F_{\rm sig}(x)~=~f_{\rm b}^{\rm raw} \times F_{\rm b}(x)~+~(1-f_{\rm b}^{\rm raw}) \times F_{\rm P}(x),
\end{equation}
where $f_{\rm b}^{\rm raw}$ is the uncorrected fraction of non-prompt \jpsi and $F_{\rm b}(x)$ and $F_{\rm P}(x)$ are the pseudoproper decay length distributions of non-prompt and prompt \jpsi, respectively. 

\begin{figure}[tb]
	\centering
  \includegraphics[width=0.485\textwidth]{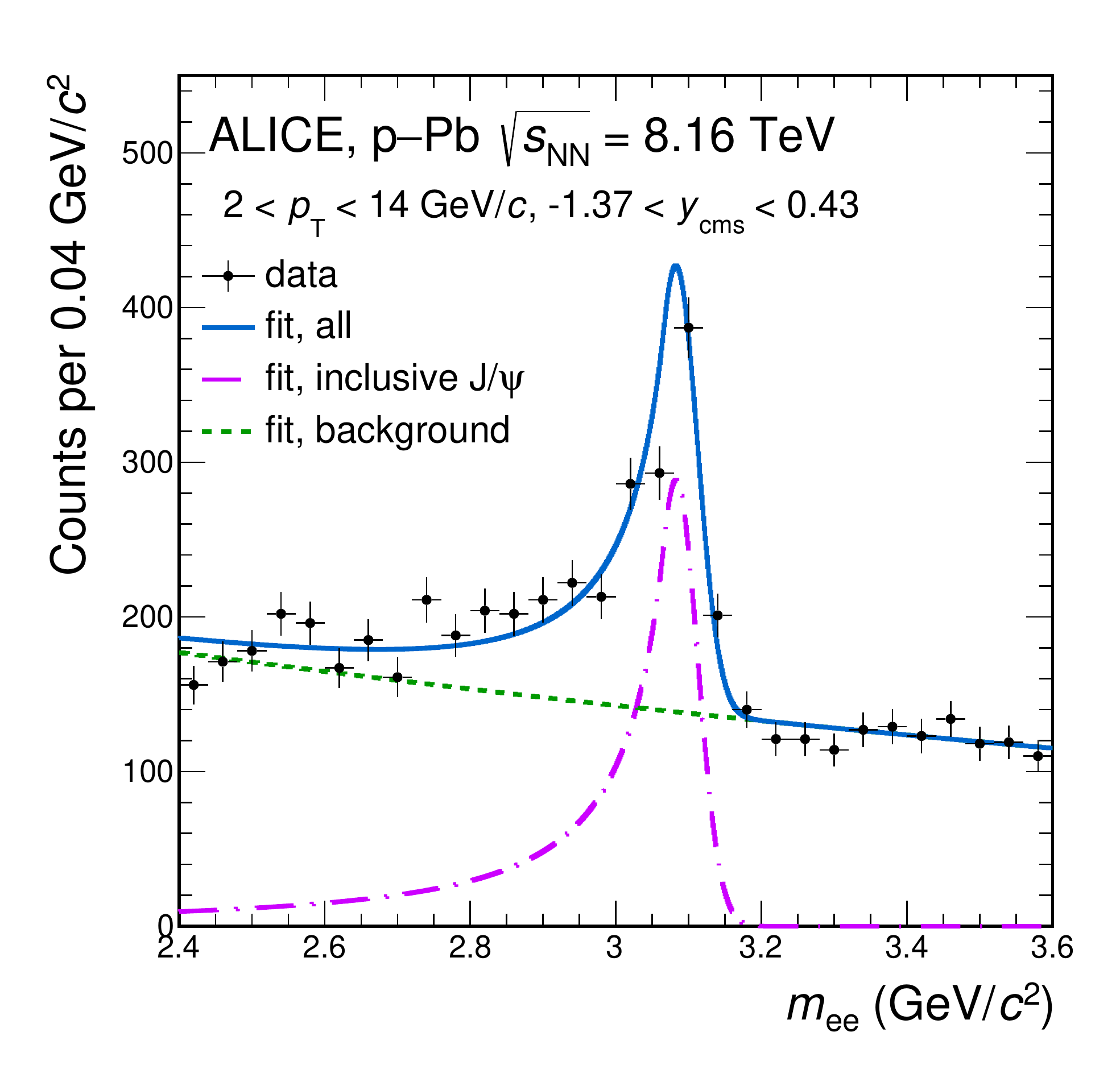}
  \includegraphics[width=0.485\textwidth]{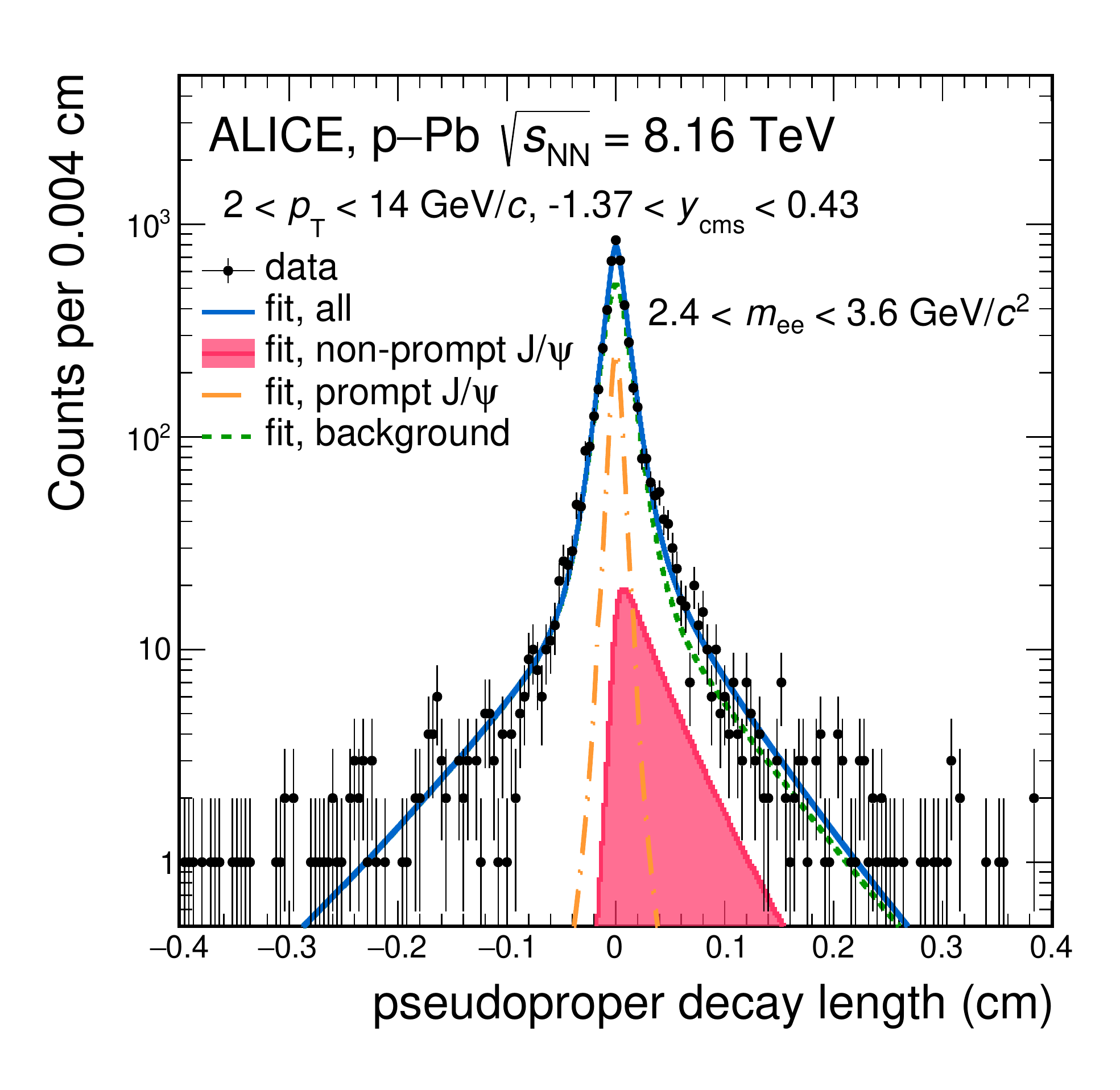}
  
	\caption{Invariant mass (left) and pseudoproper decay length (right) distributions for \ee pairs in the \pt interval $2 < \pt < 14$~\GeVc. Only statistical uncertainties are shown. The one-dimensional projections of the different contributions of the fit as described in the text are drawn for each distribution.}
	\label{Figure::Analysisnonprompt}
\end{figure}

The invariant mass and pseudoproper decay length distributions of \ee pairs are shown as an example for the \pt interval $2 < \pt < 14$~\GeVc in Fig.~\ref{Figure::Analysisnonprompt}. The one-dimensional projections of the different components of the fit are drawn for each distribution. 
The invariant mass distribution is described in the fit by a Crystal Ball function~\cite{ALICE-PUBLIC-2015-006} and an exponential function for the signal and background contributions, respectively. The parameters of the Crystal Ball function were tuned to match the \jpsi signal shape in MC, which describes well the measured invariant mass distribution of inclusive \jpsi mesons presented in this article.  
The analysis of the non-prompt \jpsi fraction was carried out for $\pt > 2$~\GeVc in coarser momentum intervals than the inclusive analysis to reduce statistical fluctuations.

The pseudoproper decay length distribution of prompt \jpsi, known as the resolution function $R(x)$, is given by the finite detector resolution and reconstruction algorithm. The resolution function was determined from MC simulations and is well described by the sum of two Gaussian distributions at its core, with the addition of a power law function for the tails, symmetric around $x = 0$. To minimise discrepancies in the $R(x)$ description between data and MC simulations, the distance-of-closest-approach (DCA) to the primary vertex in the transverse plane for single tracks was tuned in MC simulations via a data-driven approach. In this approach, differences in the mean and width of the DCA distributions  were corrected for in MC simulations as a function of \pt, azimuthal angle and SPD hit configuration, which strongly influences the DCA resolution, using primary pions to match the performance in data.  The RMS of the resolution function shows, as expected, a strong dependence on the \jpsi \pt with values of about $156~ \mu {\rm m}$ ($43~\mu {\rm m}$) for a \jpsi \pt of 2~\GeVc (12~\GeVc). The pseudoproper decay length distribution of non-prompt \jpsi is modelled by the kinematic distribution of \jpsi from b-hadron decays obtained from the MC simulation convoluted with the resolution function. The relative fractions of the different b-hadron species were reweighted in the MC simulations to match the measurements performed by the LHCb collaboration in pp collisions at $\s = 13$~TeV~\cite{LHCb_2019_bhadron}. These measurements are consistent with the results, available in a coarser \pt binning, by the LHCb collaboration in \pPb collisions at \eightnn~\cite{LHCb:2019avm}. As the background contribution changes with invariant mass, see Fig.~\ref{Figure::Analysisnonprompt} (left), $F_{\rm bkgd}(x)$ was obtained by fitting the pseudoproper decay length distributions from the side bands of the invariant mass distribution ($2.4 < m_{\rm ee} < 2.8$ ~\GeVmass and $3.2 < m_{\rm ee} < 3.6$~\GeVmass) and interpolating the background contribution under the \jpsi signal peak ($2.8 < m_{\rm ee} < 3.2$~\GeVmass) assuming a linear combination of the background in the left and right side band. The relative fraction of each contribution was included as an additional free parameter in the global fitting.

The final $f_{\rm b}$ was then obtained by correcting $f_{\rm b}^{\rm raw}$ by the average acceptance and efficiency of prompt $\left( \langle A \times \epsilon  \rangle_{\rm P} \right)$ and non-prompt $\left( \langle A \times \epsilon  \rangle_{\rm b} \right)$ \jpsi in a given \pt interval:
\begin{equation}
    f_{\rm b}~=~\left( 1 + \frac{1-f_{\rm b}^{\rm raw}}{f_{\rm b}^{\rm raw}} \times \frac{ \langle A \times \epsilon \rangle_{\rm b}}{\langle A \times \epsilon \rangle_{\rm P}} \right)^{-1}.
\end{equation}
The factors $ \langle A \times \epsilon \rangle $ obtained from the MC simulations are slightly different for prompt and non-prompt \jpsi due to the different \pt distributions. For a more realistic treatment, the \pt distributions in the MC simulations were reweighted to match measurements of prompt and non-prompt \jpsi mesons performed by the LHCb collaboration at forward rapidity in \pPb collisions at \eightnn~\cite{LHCb:2017ygo}.  

The systematic uncertainties of the $f_{\rm b}$ measurement arising from the imprecise knowledge of the probability distributions used in the two-dimensional fit as well as from  the input MC \pt distributions of prompt and non-prompt \jpsi affecting the $ \langle A \times \epsilon \rangle $ correction factors are summarised in Table~\ref{tab:Sysunc_fb}.

To estimate the systematic uncertainty related to the calculation of the average correction factors $\left( \langle A \times \epsilon \rangle \right)$, the \pt distributions in MC were not reweighted with the measurements by the LHCb collaboration, but instead with ALICE measurements at midrapidity in \pPb collisions at \fivenn~\cite{ALICE:2021lmn}. Due to the coarse binning and the large uncertainties, the central values of the ALICE measurement were fitted with a power law function, which was then used in the reweighting procedure. The differences in the \fb fractions corrected with the $ \langle A \times \epsilon \rangle $ factors obtained from different \pt shapes were then taken as systematic uncertainties. The uncertainties are largest for the \pt-integrated case as well as at low \pt, where the \pt distributions rapidly change. As in previous analyses, possible systematic uncertainties due to polarisation are not considered as the measured degree of polarisation is small in pp collisions~\cite{ALICE:2011gej,LHCb:2011zfl,CMS:2013gbz} and no measurement yet exists for \pPb collisions. In the MC simulations used, no polarisation is implemented for prompt \jpsi, while due to the contributions of the different b-hadron species a small polarisation arises for non-prompt \jpsi as implemented in EvtGen. Assuming no polarisation also for non-prompt \jpsi would lead, as studied in Ref.~\cite{ALICE:2021lmn}, to a 4\% (1\%) variation of $ \langle A \times \epsilon \rangle $ at low (high)~\pt.

\begin{table}[tb]
    \centering
     \caption{Systematic uncertainties (in percent) of the \pt-differential and \pt-integrated $f_{\rm b}$ measurements.}
    \label{tab:Sysunc_fb}
    \begin{tabular}{l|c|c|c|c|c|c}
        &  \multicolumn{6}{c}{\pt (\GeVc)} \\
    Sources  & 2$-$4 & 4$-$6 & 6$-$8 & 8$-$10 & 10$-$14 &2$-$14 \\
    \hline
    MC input \pt shape & 3.0 & 1.6 & 1.0 & 0.8 & 0.5 & 6.7 \\      
    Resolution function $R(x)$  & 3.4 & 1.0 & 0.5 & 0.4 & 0.4 & 1.3 \\   
    $x$ distr. of non-prompt \jpsi $\left( F_{\rm b}(x) \right)$ & 6.4 & 1.1 & 1.7 & 2.2 & 0.5 & 3.8 \\  
    $x$ distr. of bkgd $\left( F_{\rm bkgd}(x) \right)$ & 3.3 & 3.7 & 2.0 &1.6 & 2.1 & 3.8 \\  
    Inv. mass p.d.f. of signal $\left( M_{\rm sig}(m_{\rm ee}) \right)$  & 2.6 & 1.9 & 0.6 & 2.2 & 4.8 & 2.1 \\  
    Inv. mass p.d.f. of bkgd $\left( M_{\rm bkgd}(m_{\rm ee}) \right)$ & 1.5 & 0.6 & 0.1 & 2.4 & 0.1 & 0.6  \\  
    \hline
    Total & 9.0 & 4.7 & 2.9 & 4.3 & 5.3 & 9.0 \\
    \end{tabular}
   
\end{table}

The description of the DCA distribution for single tracks in the MC simulation was improved via a data-driven approach using charged pions to reduce the systematic uncertainties of the resolution function $R(x)$. To study the influence of bremsstrahlung for electrons, the DCA resolution was varied based on the observed resolution difference between electrons and pions in the MC simulation. Two extreme scenarios were considered: one where the bremsstrahlung effect is twice as large in data as in MC, and the other where the bremsstrahlung effect is null in data. The \fb values were extracted for both hypotheses and the differences with respect to the standard scenario were taken as the systematic uncertainty of the resolution function $R(x)$. 

The \pt-differential spectra of the b-hadron species used as input for building the pseudoproper decay length distribution $F_{\rm b}(x)$ of \ee pairs from non-prompt \jpsi decays were simulated using \linebreak PYTHIA~6.4~\cite{Sjostrand:2006za}, which yields compatible results to those obtained with FONLL~\cite{Cacciari:1998it,Cacciari:2001td,Cacciari:2012ny}. As the latter describes measurements in the b-hadron sector in pp collisions well~\cite{ALICE:2012acz,ALICE:2021mgk,ALICE:2012vpz}, no additional systematic uncertainties related to the \pt-shapes were added. 
Uncertainties related to the decay kinematics were studied using PYTHIA~6.4 instead of the event generator EvtGen and the absolute differences in the resulting \fb values were assigned as the systematic uncertainties of the functional forms of the pseudoproper decay length of non-prompt \jpsi. \\
The $F_{\rm bkgd}(x)$ uncertainties were evaluated by changing the width of the side band regions as well as the extrapolated region under the signal peak. \\
The systematic uncertainties related to the signal and background templates used for fitting the invariant mass distributions of the \ee pairs were estimated by exchanging the MC \jpsi signal shape with one including only pairs originating from radiative or non-radiative decays, and by using different background fit functions (first and second order polynomials) as well as the invariant mass distribution of the same-sign electron pairs. The usage of the MC signal shape with either only radiative or non-radiative decays leads to extreme variations of the tail of the invariant mass distribution towards lower masses. 
The differences between the \fb values resulting from the  aforementioned variations of the template and the default \fb values were taken as a systematic uncertainty. \\ 
Studies with a dedicated MC simulation showed that differences in TRD trigger efficiency for prompt and non-prompt \jpsi are negligible.

All the aforementioned uncertainties were added in quadrature yielding the total uncertainties listed in Table~\ref{tab:Sysunc_fb}.

\section{Proton$-$proton reference}\label{Section::ppRef}

In order to evaluate the impact of nuclear effects on the inclusive as well as prompt and non-prompt \jpsi production in \pPb collisions, reference cross sections are needed which reflect the production in the absence of cold nuclear matter and hot-medium related effects. Since measurements in proton-proton collisions at the same collision energy are not available, the reference distributions had to be obtained from measured data at different centre-of-mass energies. The procedures are described in the following.

\subsection{Reference for the inclusive \jpsi analysis}\label{Section::ppRefInclusive}

The calculation of the inclusive \jpsi production cross section at $\sqrt{s}~=~8.16$~TeV is based on an assumption on the shape of the \jpsi \pt-differential cross section as well as an interpolation between measured results at different collision energies to obtain the \pt-integrated cross section and the average transverse momentum $\left( \MeanPt \right)$. With a suitable transformation, the \jpsi \pt-differential cross section can be described by a universal function~\cite{Bossu:2011qe}, independent of collision energy and rapidity. The universal function is defined as
\begin{equation}
    \frac{\MeanPt}{\der \sigma/\der y} \times \frac{\der ^2\sigma}{\der y \der \pt} =
    \frac{2(n-1) \cdot C^2 \times \pt/\MeanPt}{
    (1 + C^2 \times (\pt/\MeanPt)^2)^n},
    \label{eq:PtParametr}
\end{equation}
with $C = \Gamma(3/2)\Gamma(n-3/2)/\Gamma(n-1)$, where $n$ is left as the only free fit parameter if the values of the \pt-integrated \jpsi cross section and \MeanPt are known. The universal function was fitted to the available \jpsi data~\cite{Acosta:2004yw,CMS:2010nis,ATLAS:2011aqv,LHCb:2011zfl,LHCb:2013itw,ALICE:2017leg,CMS:2017exb,LHCb:2015foc,ALICE:2021dtt}, which range from 1.96 to 13~TeV in centre-of-mass energy with a \pt range from zero to 20~\GeVc, resulting in $n = 3.45 \pm 0.05$. The uncertainty of the parameter $n$ was obtained by excluding individual \jpsi measurements and repeating the fit. 
Once the $\MeanPt$ and the \pt-integrated cross section are known, the \pt-differential reference at $\s = 8.16$~TeV can then be calculated.

The $\MeanPt$ value of $ 2.86 \pm 0.03 $~\GeVc at $\s = 8.16$~TeV was evaluated from the fit function $\MeanPt (\s) = a + b \times \log(\s)$, which describes the evolution of the measured \MeanPt at midrapidity well over three orders of magnitude as a function of collision energy from 0.2 to 13~TeV~\cite{ALICE:2019pid}.

For the interpolation of the \pt-integrated \jpsi cross section at midrapidity, it is assumed that the collision-energy dependence of \jpsi production is the same as the one of $\ccBar$ quark pair production. As FONLL describes the available measurements of $\sigma_{\ccBar}$ well ~\cite{ALICE:2021dhb}, the \pt-integrated \jpsi production cross section was estimated by scaling the $\ccBar$ cross sections calculated by FONLL with the PDF set CTEQ $6.6$~\cite{Cacciari:1998it,Cacciari:2001td,Cacciari:2012ny}. The $\ccBar$  cross section $\sigma_{\ccBar} (|y|<0.5) = \der \sigma_{\ccBar}/\der y$ was calculated for midrapidity and for $\pt < 50$~\GeVc for all collision energies from 0.2 to 13~TeV for which measurements of the \pt-integrated \jpsi production cross section at midrapidity and low \pt exist~\cite{Acosta:2004yw,Adare:2006kf,Aamodt:2011gj,Abelev:2012kr,ALICE:2019pid,ALICE:2021dtt}. Via a $\chi^2$ minimisation, the collision energy dependence of the FONLL cross sections was scaled to the measured \jpsi production cross sections showing a good agreement between the energy dependence of the FONLL calculations and the one of the measured data. The model uncertainties, i.e.\ the charm quark mass, renormalisation and factorisation scales, are large, but were assumed to be fully correlated over the collision energy and thus should not change the shape of the energy dependence.
The cross section was then extracted by evaluating the scaled FONLL curve at $\s = 8.16$~TeV. For the estimation of the statistical and systematic uncertainties of the scaling procedure, each data point was shifted by its statistical or systematic uncertainty, respectively, independently from the other data points, i.e.\ the uncertainties were assumed to be uncorrelated across collision energies.
These variations were repeated multiple times leading to a Gaussian distribution with the mean being the evaluated central \jpsi cross section value and the width providing a $1\sigma$ uncertainty. The resulting \pt-integrated \jpsi cross section at $\s = 8.16$~TeV and midrapidity ($|y|~<~0.9$) is BR $\times \frac{\der \sigma}{\der y}$ = 452.3 $\pm$ 2.1 (stat.) $\pm$ 16.5 (syst.) ${\rm nb}$.

Using the universal function and the parameters obtained above, the \pt-differential \jpsi production cross section was calculated for the same \pt intervals as the \pPb measurement. The resulting pp reference is shown in Fig.~\ref{Figure::Resultspt} (left) scaled by the Pb mass number ($A$~=~208).
The systematic uncertainty has a correlated and uncorrelated component across \pt intervals. The correlated uncertainty originates from the interpolated \pt-integrated \jpsi cross section and is about 3.8\%. The uncorrelated uncertainty is composed of the statistical and systematic uncertainty of the input \pt spectra and \MeanPt as well as the shape uncertainty of the universal fit function. The latter is the dominating contribution to the uncertainty and was estimated to be 1\% for $\pt < 7$ \GeVc and 5\% above. It was determined by comparing the \pt shape obtained with the universal fit function to the measured distributions for several collision energies. For illustration purposes, the correlated and uncorrelated uncertainties were added in quadrature and are shown as a grey band in Fig.~\ref{Figure::Resultspt} (left).

\subsection{Reference for the non-prompt \jpsi fraction analysis}\label{subsectionpprefnonprompt}
The non-prompt \jpsi fraction \fb as a function of \pt in pp collisions shows a slight dependence on centre-of-mass energy. As no measured pp reference at $\s = 8.16$~TeV exists, the \pt-differential  \fb fractions were obtained via an interpolation method based on measurements available at midrapidity at other centre-of-mass energies ranging from 1.96~TeV to 13~TeV~\cite{Acosta:2004yw, ALICE:2021edd, CMS:2017exb, CMS:2010nis, ATLASb2jpsi_pp78TeV}. At each energy with available measurements, the \pt-differential \fb values with their statistical and systematic uncertainties added in quadrature were fitted with a function. The latter was obtained as the ratio of the FONLL calculated \pt-differential production cross section of non-prompt \jpsi and the universal function representing the inclusive \jpsi measurement (see Sec.~\ref{Section::ppRefInclusive}). The parameters of the universal fit function were not constrained, to allow for the varying \pt-dependence of \fb measured at each collision energy.  For the fit results at each energy a 1$\sigma$ uncertainty band was obtained based on the experimental uncertainties of the considered measurements as well as the FONLL uncertainties (beauty quark mass, renormalisation and factorisation scale), with the latter being the dominant contribution.

In the next step, the \fb fractions evaluated in narrow \pt intervals from the fitted functions were parameterised as a function of centre-of-mass energy assuming a linear dependence to obtain the \fb value for $\s = 8.16$~TeV. Repeating this procedure with the upper and lower edge of the 1$\sigma$ uncertainty band of the fit function yielded the systematic uncertainties of the \fb values used as pp reference. Replacing the linear function by a power law or exponential function resulted in a negligible systematic uncertainty. 

As the \pt intervals used in this analysis are coarse and the \fb values change within the \pt intervals, the average \fb values and their related uncertainties were then determined for each \pt interval using the \pt distribution as a weight. The uncertainty amounts to 6\% in the first \pt interval, $2 < \pt < 4$~\GeVc, and is negligible above.

\section{Results}\label{Section::Results} 

\begin{figure}[tb]

  \includegraphics[width=0.54\textwidth]{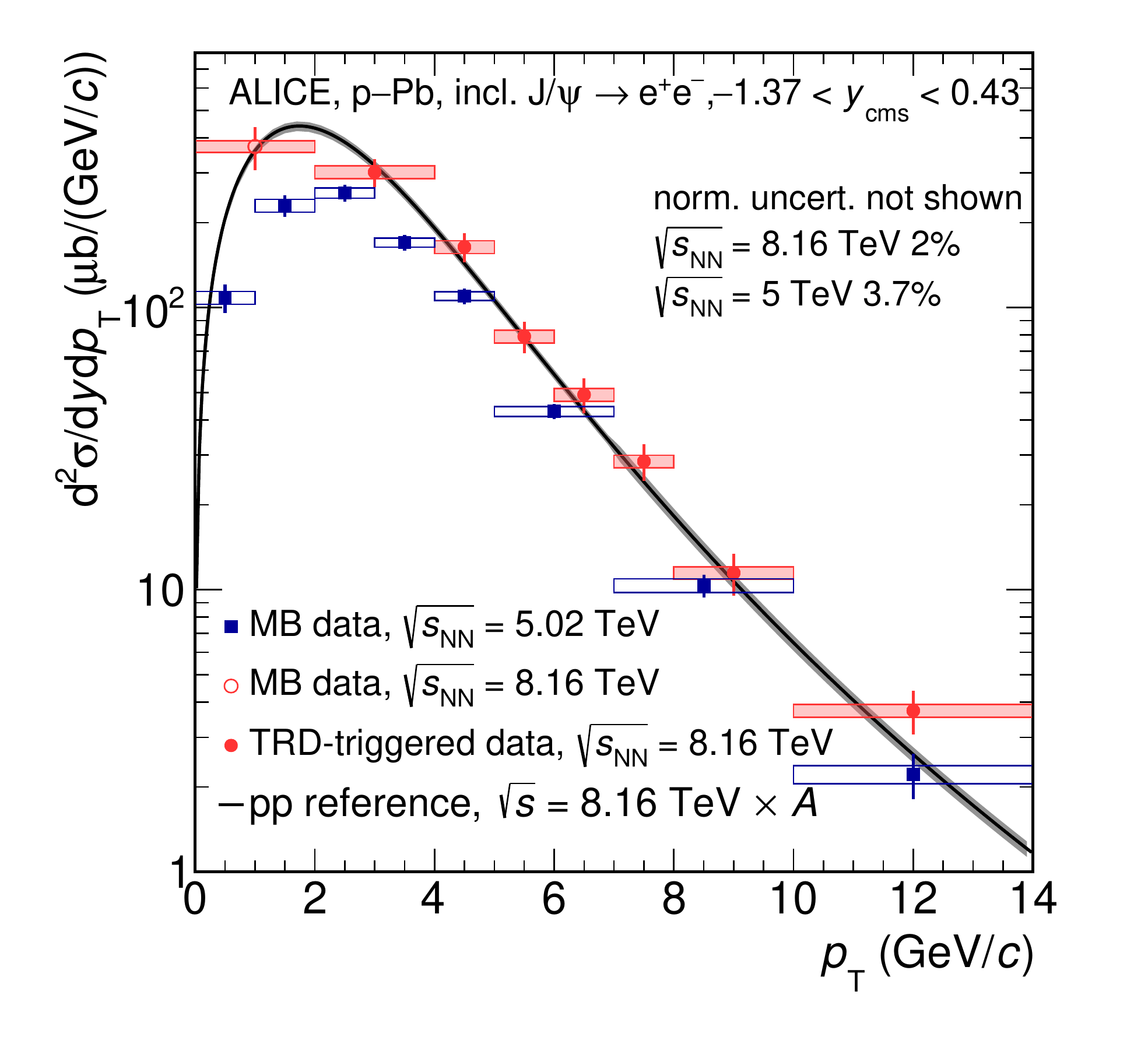}
  \hspace{-0.7cm} 
   \includegraphics[width=0.54\textwidth]{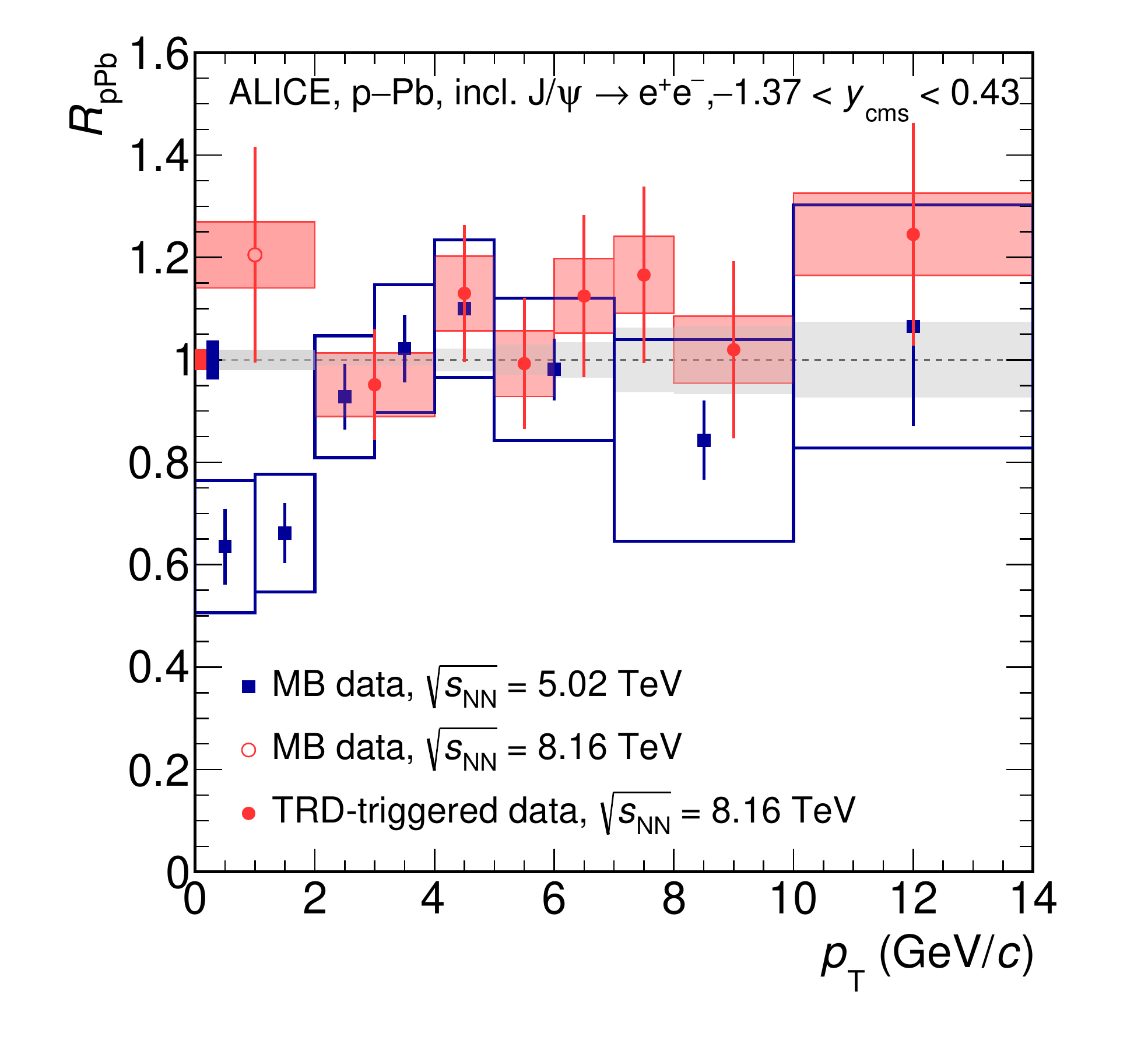}
   
	\caption{(left) \pt-differential inclusive \jpsi production cross section in MB and TRD-triggered \pPb collisions at \eightnn in comparison with the measurements at \fivenn~\cite{ALICE:2021lmn}. The statistical and systematic uncertainties are shown as error bars and boxes. Also depicted is the pp reference spectrum at $\sqrt{s} = 8.16$~TeV obtained via the procedure described in Sec.~\ref{Section::ppRef} and multiplied by the mass number of the Pb nucleus ($A = 208$). The grey band represents the combined correlated and uncorrelated uncertainties. (right) \RpPb of inclusive \jpsi in MB and TRD-triggered \pPb collisions at \fivenn and 8.16~TeV. The statistical and systematic uncertainties are shown as error bars and boxes. For the 8.16~TeV measurement, the uncorrelated uncertainties of the pp reference are shown as a band around unity. The normalisation uncertainties are shown as coloured boxes around unity at zero \pt.  }
	\label{Figure::Resultspt}
\end{figure}

The \pt-differential production cross section of inclusive \jpsi in \pPb collisions at \eightnn is shown in Fig.~\ref{Figure::Resultspt} (left). The statistical and systematic uncertainties are shown as bars and boxes, respectively. The measurements cover the \pt intervals 0$-$2~\GeVc and from 2 up to 14~\GeVc obtained from the MB and the TRD-triggered data samples, respectively. The usage of the single-electron triggers provided by the TRD strongly enhances the number of \jpsi measured at intermediate and high \pt.  

As expected for the higher collision energy, the \pt-differential production cross section at \linebreak \eightnn is consistently above the measurement at \fivenn, albeit the shapes are very similar for $\pt > 2$~\GeVc.  Also depicted in Fig.~\ref{Figure::Resultspt} (left) is the pp reference spectrum for $\sqrt{s} = 8.16$~TeV obtained as discussed in Sec.~\ref{Section::ppRef} and scaled by the mass number of Pb ($A = 208$). 
It can be observed that the modifications in \pPb compared to pp collisions are small for $\pt > 2$~\GeVc.       

To quantify the nuclear effects, the nuclear modification factor \RpPb is calculated from the production cross sections in \pPb and pp collisions:
\begin{equation}
    \RpPb~=~\frac{\der^2 \sigma_{{\rm J}/\psi}^{\rm pPb} /\der y \der \pt}{A \times \der^2 \sigma_{{\rm J}/\psi}^{\rm pp} / \der y \der \pt},
    \label{equation:RpPb}
\end{equation}
where $A$ is the mass number of the Pb nucleus and $\der^2 \sigma_{{\rm J}/\psi}^{\rm pPb} /\der y \der \pt$ and $\der^2 \sigma_{{\rm J}/\psi}^{\rm pp} / \der y \der \pt$ are the \pt-differential production cross sections in \pPb and pp collisions at the same collision energy. The nuclear modification factor is expected to be equal to unity in the absence of nuclear effects.

Figure~\ref{Figure::Resultspt} (right) shows the \pt-differential \RpPb values for inclusive \jpsi production. The error bars represent the statistical uncertainties, the boxes the systematic uncertainties. The latter were derived by adding in quadrature the systematic uncertainties of the \pPb result and the systematic uncertainties correlated in $\pt$ of the pp reference spectrum. The uncorrelated uncertainties of the pp reference spectrum are shown as a band around unity. The normalisation uncertainty is shown as a box around unity at zero \pt. 

For $\pt > 2$~\GeVc, the \RpPb is consistent with unity. This tendency was already observed at the lower energy of \fivenn. The data indicate that nuclear effects do not exceed 20\% in the kinematic range above $\pt~=~2$~\GeVc taking into account the deviation from unity and the related statistical and systematic uncertainties. The \RpPb for $\pt < 2$~\GeVc from the MB analysis is, considering uncertainties, larger than the one at the lower centre-of-mass energy by two standard deviations. However, no significant conclusion can be drawn due to the limited size of the MB data sample.

To enhance the precision of the \pt-integrated \RpPb value, an extrapolation down to zero-\pt of the cross section measured with the TRD-triggered data sample is used instead of the minimum bias result. The measured visible cross section for $\pt > 2$~\GeVc corresponds to about two-thirds of the \pt-integrated cross section at midrapidity. The extrapolation method uses the \pt-differential measurement in \pPb collisions at \fivenn assuming that cold nuclear matter effects influencing \jpsi production are the same at both collision energies. This is very well supported by the findings at forward and backward rapidity by the ALICE and LHCb collaborations within their present measurement accuracy~\cite{ALICE:2018mml,LHCb:2017ygo}. The extrapolation factor~$F$ was computed as the ratio of the cross section in the \pt~interval 0$-$2~\GeVc to the one in 2$-$14~\GeVc. Taking into account the hardening of the \pt~spectrum as observed in pp collisions between the two collision energies as estimated using the universal function (see Sec.~\ref{Section::ppRef}) with the mean~\pt parameters for each energy, the extrapolation factor $F = 0.46 \pm 0.04 {\rm (stat.)} \pm 0.04 {\rm (syst.)}$ was derived. The statistical and systematic uncertainties of the extrapolation factor were determined by adding in quadrature or linearly, respectively, the uncertainties of each \pt interval of the \pPb 5.02~TeV cross section measurement. The systematic uncertainty also includes the uncertainties related to the evaluation of the hardening of the \pt~spectrum and to possible differences in cold nuclear matter effects at both energies. The first contribution was determined by modifying the parameters of the universal function by the uncertainties of the mean~\pt values for both energies, resulting in a modification of the $F$~factor by~3\%. The second contribution was estimated by computing the difference in the \jpsi cross section calculations including cold nuclear matter effects via the EPPS16 nPDF sets by Lansberg $et~al.$~\cite{Lansberg:2016deg,Shao:2015vga} at both energies, assuming the uncertainties of the calculations to be highly correlated between energies. Repeating the calculation of the $F$~factor with the obtained difference yields a 2.5\% uncertainty. All contributions to the total systematic uncertainty of the extrapolation factor~$F$ were added quadratically.

\begin{figure}[tb]
	\centering
  \includegraphics[width=0.6\textwidth]{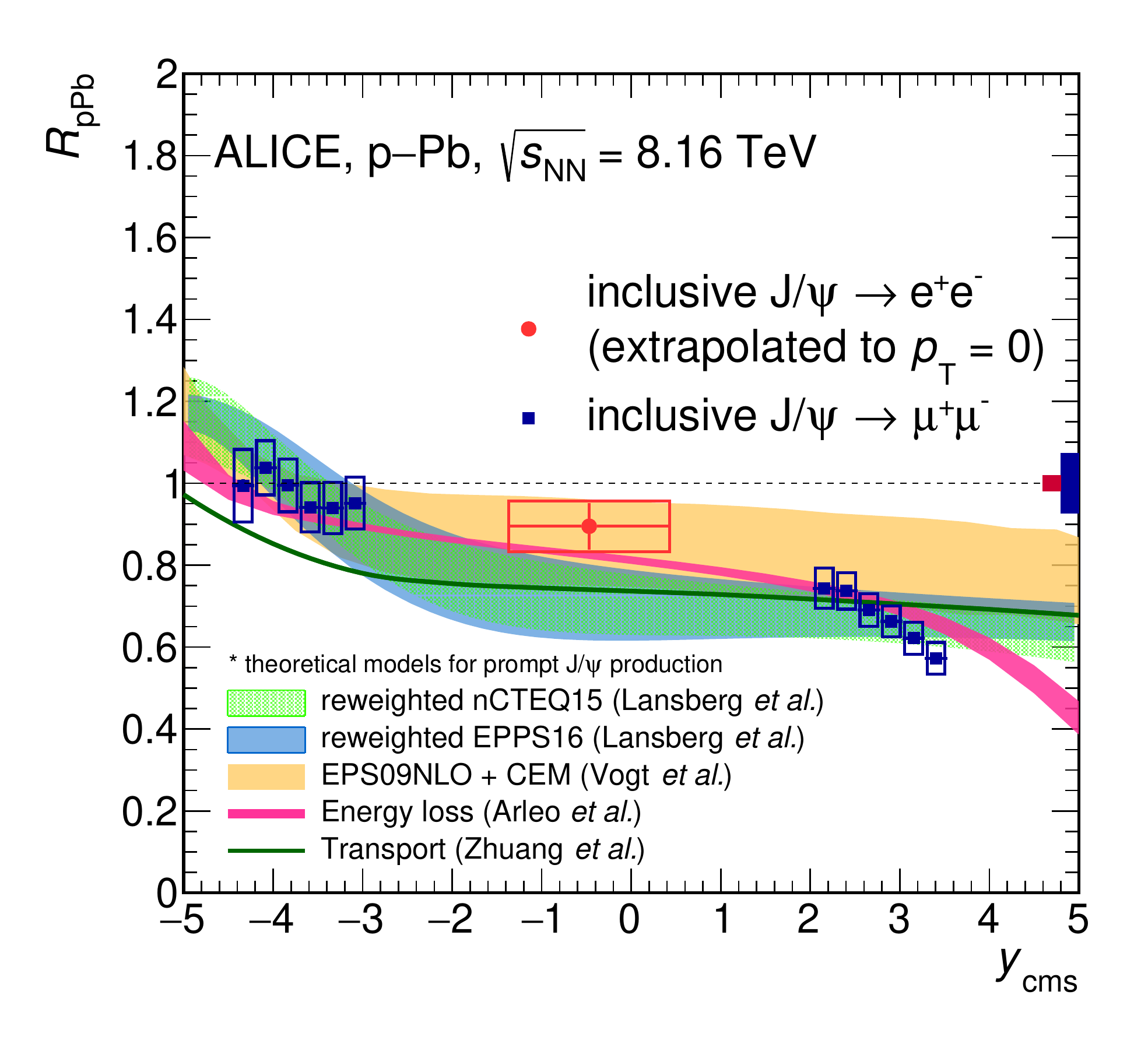}
	\caption{\pt-integrated \RpPb values for inclusive \jpsi production as a function of $y_{\rm cms}$ in comparison with results at forward and backward rapidity by the ALICE collaboration~\cite{ALICE:2018mml}. The statistical and systematic uncertainties are shown as error bars and boxes. The overall normalisation uncertainties are shown as boxes around unity at large $y_{\rm cms}$. Also shown are the results of theoretical calculations which refer to prompt \jpsi ~\cite{Chen:2016dke,Arleo:2013zua,Albacete:2017qng,Lansberg:2016deg,Kusina:2017gkz, EPPS16_2017, nCTEQ15_2016}.}
	\label{Figure::Resultsy}
\end{figure}

The extrapolated \pt-integrated inclusive \jpsi cross section is then given by:
\begin{equation}
    \frac{{\rm d}\sigma_{\rm inclusive~J/\psi}^{\rm extra}}{{\rm d}y}~=~ (1~+~F) \times \frac{{\rm d}\sigma_{\rm inclusive~J/\psi}^{\rm vis}}{{\rm d}y} = 1409 \pm 89 {\rm (stat.)} \pm  84 {\rm (syst.)}  \mu{\rm b},
\end{equation}
where $\frac{{\rm d}\sigma_{\rm inclusive~J/\psi}^{\rm vis}}{{\rm d}y} = 968 \pm 56 {\rm (stat.)} \pm  50 {\rm (syst.)} \mu{\rm b}$ denotes the measured visible cross section of inclusive \jpsi in the \pt interval $2 < \pt < 14$~\GeVc at midrapidity ($-1.37 < y_{\rm cms} < 0.43$) in \pPb collisions at \eightnn. The \pt-integrated \RpPb value, depicted in Fig.~\ref{Figure::Resultsy}, was then obtained as in Eq.~\ref{equation:RpPb}. The \pt-integrated cross section for pp collisions at the same collision energy was obtained via the interpolation method discussed in Sec.~\ref{Section::ppRef}. The systematic uncertainties of the \RpPb value were calculated by adding in quadrature the systematic uncertainties of the extrapolated \pt-integrated inclusive \jpsi cross section in \pPb and of the cross section in pp collisions.

Also shown in Fig.~\ref{Figure::Resultsy} are the ALICE measurements for inclusive \jpsi in the dimuon decay channel at forward $(2.03 < y_{\rm cms} < 3.53)$ and backward $(-4.46 < y_{\rm cms} < -2.96)$ rapidity~\cite{ALICE:2018mml} at the same collision energy, as well as several theoretical calculations for prompt \jpsi. The calculations by Lansberg $et~al.$ are based on the framework of NRQCD factorisation with nCTEQ15 and EPPS16 nPDF sets that were reweighted to include results from the RHIC and LHC colliders~\cite{Lansberg:2016deg,Shao:2015vga,Eskola:2016oht}. The calculation by Vogt $et~al.$ is based on a pure shadowing scenario employing the next-to-leading order (NLO) Color Evaporation Model (CEM) with the EPS09 shadowing parametrisation~\cite{Albacete:2013ei}. This older parton distribution function was obtained before collider data were available. The calculation by Arleo $et~al.$~\cite{Arleo:2014oha} includes effects of momentum broadening, coherent parton energy loss and no nuclear shadowing of the gluon PDF. The model by Zhuang $et~al.$ ~\cite{Chen:2016dke} includes final-state effects (so-called hot nuclear matter effects), where the \ccBar states interact with the system generated in the collision, as well as nuclear shadowing using the EPS09 gluon nPDF. 
The theoretical models describing the forward and backward rapidity results, also agree within uncertainties with the measurement at midrapidity. While the \pt-integrated inclusive yield is strongly dominated by the prompt \jpsi, the \pt-differential results for prompt and non-prompt \jpsi are separately compared with models, in the following part of this section.

\begin{figure}[tb]
	\centering
  \includegraphics[width=0.6\textwidth]{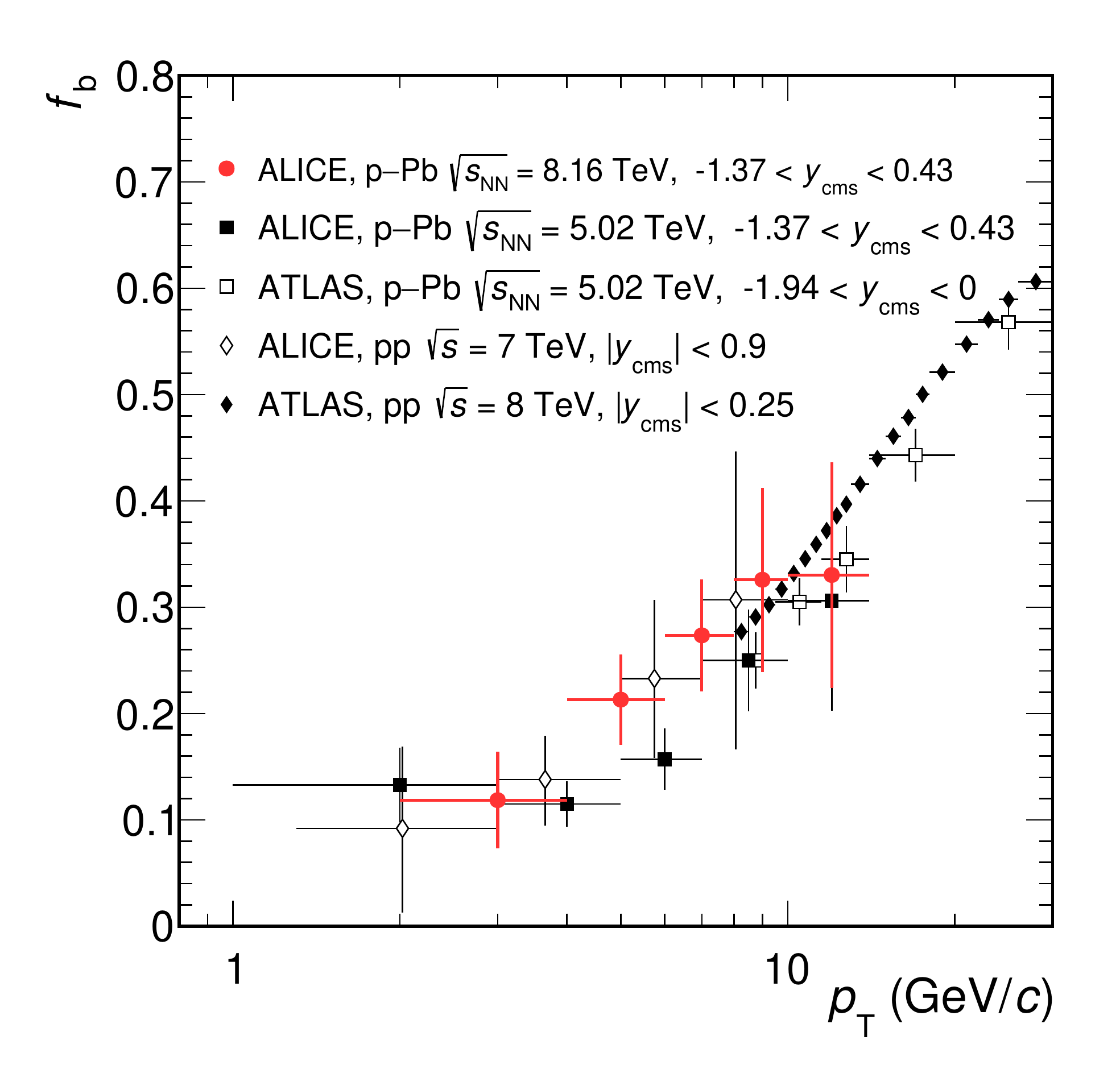}

	\caption{Fraction of \jpsi originating from $b$-hadron decays as a function of \jpsi \pt in comparison with analogous measurements, also at midrapidity, in pp and \pPb collisions by the ALICE and ATLAS collaborations~\cite{ALICE:2021lmn,ATLASb2jpsi_pPb5TeV,ALICE:2012vpz,ATLASb2jpsi_pp78TeV}. The statistical and systematic uncertainties of the measurements were added in quadrature.}
	\label{Figure::Resultsfb}
\end{figure}

The fraction of \jpsi originating from $b$-hadron decays (\fb) is shown as a function of \jpsi \pt in Fig.~\ref{Figure::Resultsfb}. In addition to the results presented in this article, similar measurements from \pPb collisions at \linebreak \fivenn and from pp collisions at 7 and 8~TeV, performed by the ALICE and ATLAS collaborations, are depicted. For all data shown, the statistical and systematic uncertainties were added quadratically. The measurements show the complementarity of the experiments at the LHC; in particular ALICE provides measurements at low and intermediate~\pt while ATLAS has results at high~\pt. In the common~\pt interval the results from the different experiments are in good agreement. The emerging picture is a rise of \fb with increasing \pt from values close to 0.1 in the 1$-$2~\GeVc interval to values exceeding 0.5 at \pt larger than 20~\GeVc. 
With the current experimental uncertainties, one cannot discern differences between the results for pp and \pPb collisions or at different collision energies. The experimental precision at low and intermediate \pt is expected to significantly improve with the much larger data samples and tracking precision of the upgraded ALICE detector in LHC Runs~3 and~4~\cite{Citron:2018lsq,ALICE:2012dtf}. 

The \pt-integrated fraction of non-prompt \jpsi for the \pt interval $2 < \pt < 14$~\GeVc at midrapidity ($-1.37 < y_{\rm cms} < 0.43$) in \pPb collisions at \eightnn is $f_{\rm b}^{\rm vis} = 0.18 \pm 0.03 {\rm (stat.)} \pm 0.02 {\rm (sys.)} $. Using the integrated inclusive cross section in the same kinematic region, the prompt and non-prompt \jpsi cross sections per unit of rapidity were obtained as:  
\begin{equation}
   \begin{split}
    \frac{{\rm d}\sigma_{\rm prompt~J/\psi}^{\rm vis}}{{\rm d}y}&~=~ (1- f_{\rm b}^{\rm vis}) \times \frac{{\rm d}\sigma_{\rm inclusive~J/\psi}^{\rm vis}}{{\rm d}y} = 797\pm 66 {\rm (stat.)} \pm  32 {\rm (syst.)} \mu{\rm b}~~ {\rm and} \\
      \frac{{\rm d}\sigma_{\rm non {\text-}prompt~J/\psi}^{\rm vis}}{{\rm d}y}&~=~ f_{\rm b}^{\rm vis} \times \frac{{\rm d}\sigma_{\rm inclusive~J/\psi}^{\rm vis}}{{\rm d}y} = 169 \pm 30 {\rm (stat.)} \pm  17  {\rm (syst.)} \mu{\rm b}.
      \end{split}
\end{equation}

\begin{figure}[tb]
	\centering
  \includegraphics[width=0.49\textwidth]{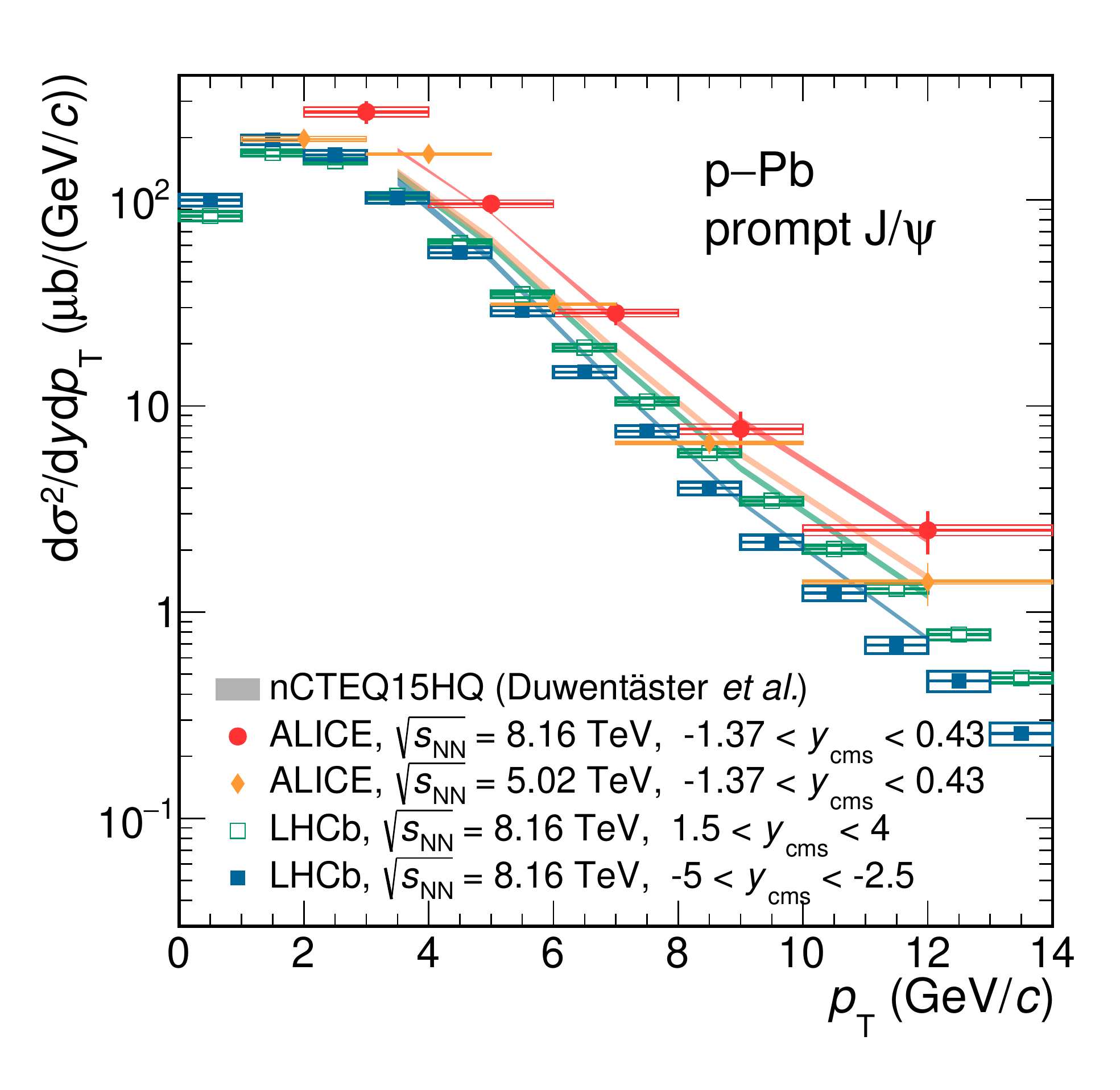}
  \includegraphics[width=0.49\textwidth]{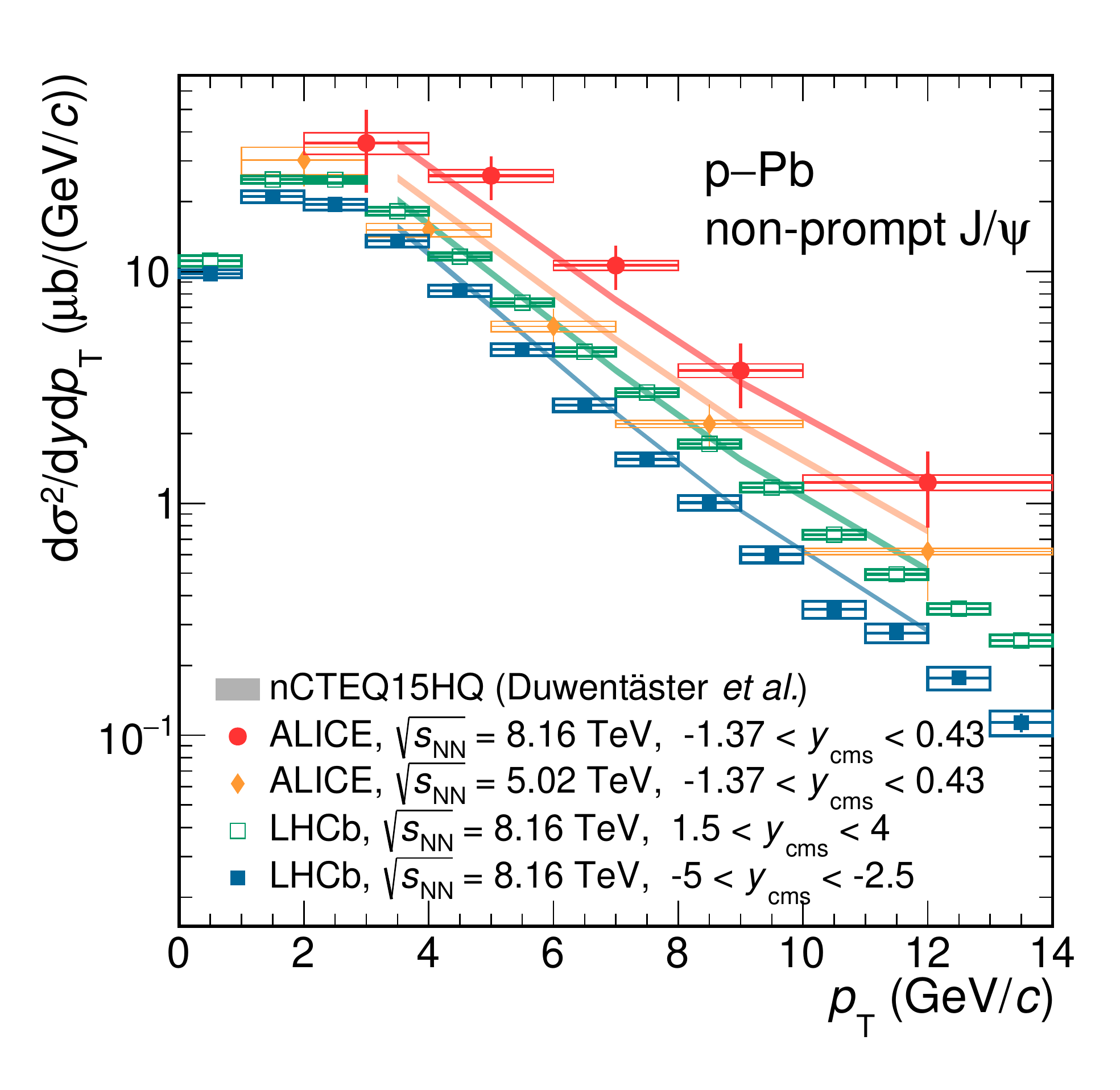}

	\caption{\pt-differential production cross sections of prompt (left) and non-prompt (right) \jpsi together with corresponding results at forward and backward rapidity by the LHCb collaboration~\cite{LHCb:2017ygo} and ALICE measurements at midrapidity at \fivenn~\cite{ALICE:2021lmn}. The statistical and systematic uncertainties are represented as error bars and boxes for each data point. The normalisation uncertainties are not shown. The measurements are compared with calculations using the latest version of the CTEQ nPDFs, nCTEQ15HQ~\cite{Duwentaster:2022kpv}, see text for details.} 
	\label{Figure::ResultspT_prompt_nonprompt}
\end{figure}

The \pt-differential production cross sections of prompt and non-prompt \jpsi are displayed in the left- and right-hand panels of Fig.~\ref{Figure::ResultspT_prompt_nonprompt} together with the corresponding measurements at forward and backward rapidity released by the LHCb collaboration at the same centre-of-mass energy, and with the ALICE measurements at midrapidity at \fivenn.

The cross section is larger at midrapidity compared to the corresponding measurements at forward and backward rapidity at the same centre-of-mass energy and shows a similar \pt dependence as the measurement at \fivenn for \pt $> 2$~\GeVc. The measurements are very well described by the calculations from P. Duwent\"aster $et~al.$ which utilise the latest version of the CTEQ15 nPDF set, the nCTEQ15HQ set~\cite{Duwentaster:2022kpv}. Here, for the cross section calculations, the effective scattering matrix elements were determined from measurements in pp collisions in a data-driven approach following Ref.~\cite{Kom:2011bd} and validated with NLO calculations in NRQCD for quarkonium and the general-mass variable-flavor-number scheme~\cite{Kniehl:2004fy} for the open heavy-flavour mesons.
To obtain this nPDF set, measurements from the LHC on heavy quark and quarkonium production were used to constrain the gluon density down to Bjorken-$x$~$\sim 10^{-5}$. Compared to the nCTEQ15 fit, not taking into account the LHC data, the uncertainty of the gluon PDF in Pb is reduced by a factor 3 around Bjorken-$x = 10^{-4}$. The measured prompt \jpsi spectrum at midrapidity is mostly driven by the nuclear gluon PDF in the Bjorken-$x$ range $2 \times 10^{-4}$ to $10^{-3}$, where the ratio of the nuclear relative to the proton gluon PDF is 0.69$-$0.79 in the new nCTEQ15HQ fit. 

The prompt and non-prompt \jpsi nuclear modification factors were calculated as
\begin{equation}
   \begin{split}
   \RpPb^{\rm prompt~J/\psi}~=~\frac{1-f_{\rm b}^{\rm pPb}}{1-f_{\rm b}^{\rm pp}} \times \RpPb ~~ {\rm and}  \\ 
   \RpPb^{\rm non{\text -}prompt~J/\psi}~=~\frac{f_{\rm b}^{\rm pPb}}{f_{\rm b}^{\rm pp}} \times \RpPb,
\end{split}
\end{equation}
where \RpPb denotes the inclusive \jpsi nuclear modification factors shown in Fig.~\ref{Figure::Resultspt}. The fractions of \jpsi originating from $b$-hadron decays in \pPb and pp collisions are represented by $f_{\rm b}^{\rm pPb}$ and $f_{\rm b}^{\rm pp}$, with the latter obtained via an interpolation procedure as discussed in Sec.~\ref{subsectionpprefnonprompt}. 

\begin{figure}[tb]
	\centering
      \includegraphics[width=0.49\textwidth]{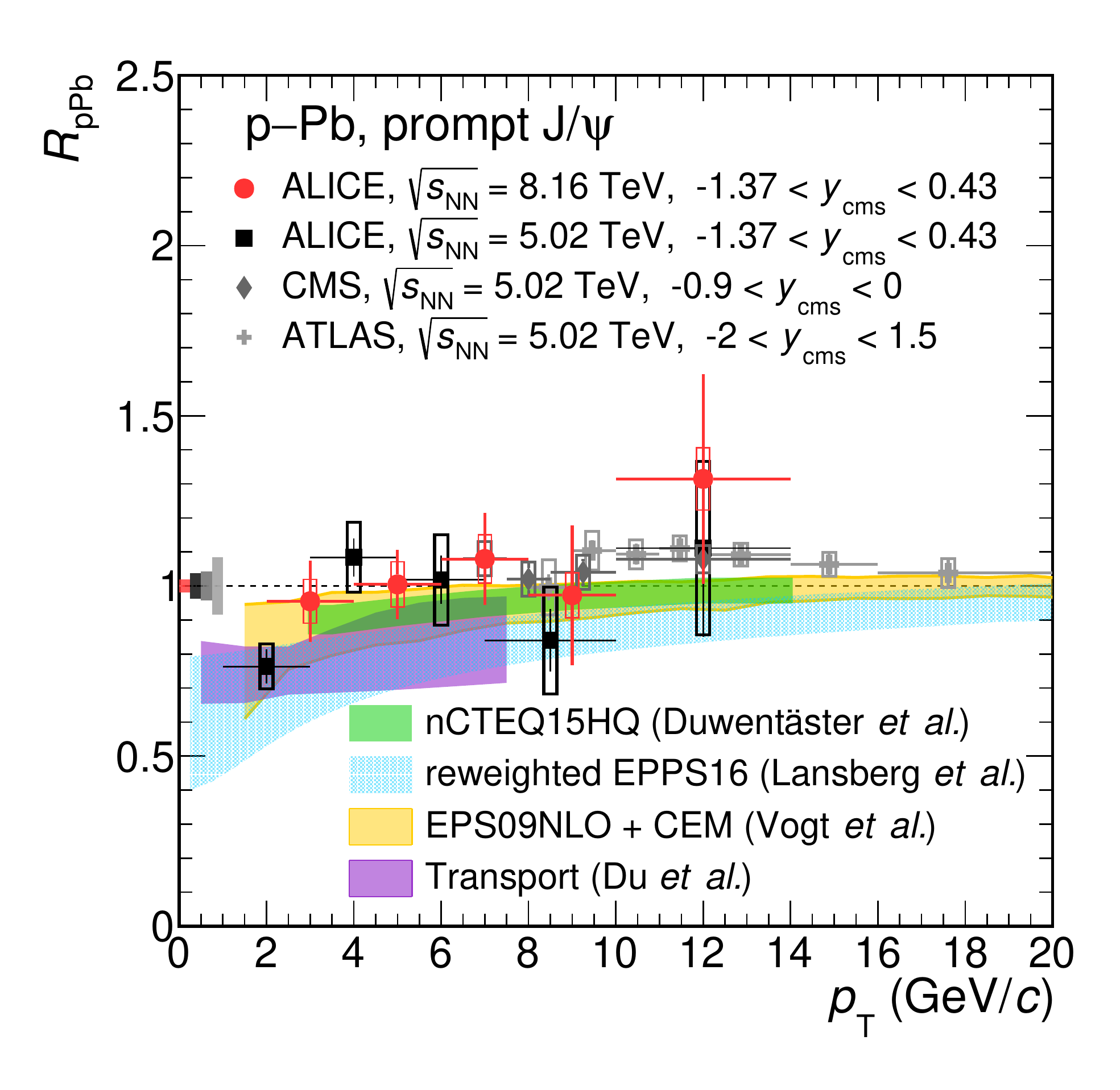}
  \includegraphics[width=0.49\textwidth]{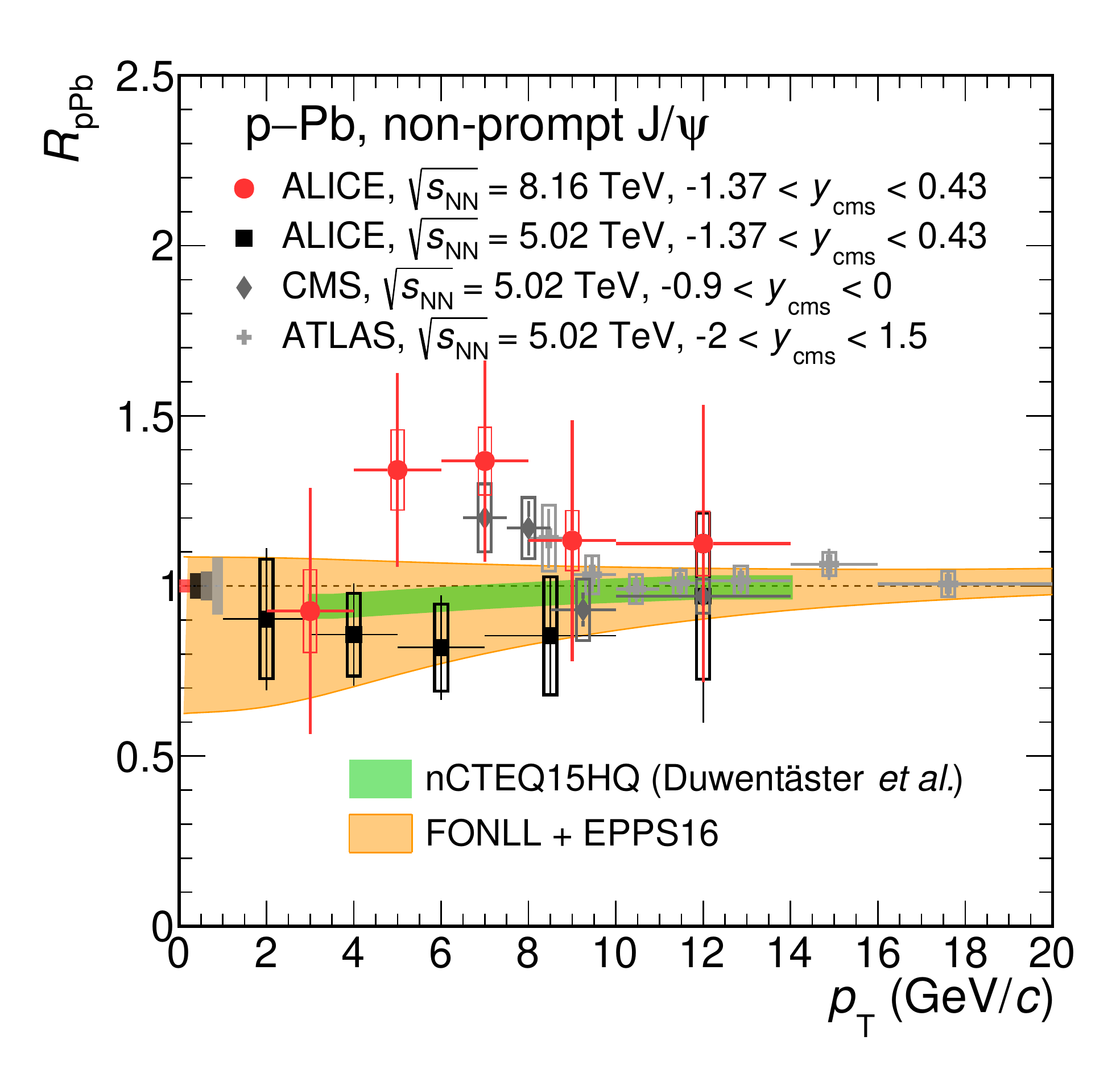}
	\caption{\pt-differential \RpPb of prompt (left) and non-prompt (right) \jpsi together with corresponding results in \pPb collisions at \fivenn~\cite{ALICE:2021lmn,CMS:2017exb,ATLASb2jpsi_pPb5TeV} and theoretical calculations~\cite{Lansberg:2016deg,Kusina:2017gkz, EPPS16_2017,Duwentaster:2022kpv,Du:2018wsj}. The statistical and systematic uncertainties are represented as error bars and boxes for each data point. The normalisation uncertainties are indicated as boxes around unity at zero \pt.}
	\label{Figure::ResultsRpPb_prompt_nonprompt}
\end{figure}

The resulting nuclear modification factors are shown as a function of \jpsi \pt in Fig.~\ref{Figure::ResultsRpPb_prompt_nonprompt} for $\pt > 2$~\GeVc. The displayed systematic uncertainties, shown as open boxes, include the uncertainties of the \fb values for pp and \pPb collisions as well as the correlated and uncorrelated uncertainties of the inclusive \jpsi \RpPb values. The prompt and non-prompt \jpsi nuclear modification factors are consistent with unity within statistical and systematic uncertainties. Measurements performed at midrapidity in \pPb collisions at \fivenn by the ALICE, ATLAS and CMS collaborations, shown for comparison, agree with our results for \pt above 2~\GeVc, while the value for the lowest \pt bin for prompt \jpsi at \fivenn from ALICE is significantly below unity. 

Several theoretical calculations for prompt \jpsi, introduced before when discussing the rapidity dependence in Fig.~\ref{Figure::Resultsy}, are shown for comparison with the measured \pt-differential \RpPb in Fig.~\ref{Figure::ResultsRpPb_prompt_nonprompt} (left).  
Also depicted is the transport calculation by Du $et~al.$~\cite{Du:2018wsj}, based on the kinetic rate-equation approach within a fireball model and previously used for heavy-ion and d$-$Au collisions, for the \pPb collision system. Shadowing effects were considered in the calculation by Vogt $et~al.$ via the pre-LHC EPS09 nPDFs. The theoretical calculations describe the low \pt data where the model should be applicable. The calculation by Lansberg $et~al.$~\cite{Lansberg:2016deg} tends to be systematically below the data in the \pt range below 10~\GeVc, only approaching unity at 20~\GeVc.
The approach based on the CEM and EPS09 nPDFs as well as the NLO pQCD calculation of Duwent\"aser $et~al.$ are closest to the data. The latter calculation has the smallest uncertainties profiting from the latest version of the nCTEQ15HQ nPDF set. This is also the case for the corresponding studies on non-prompt \jpsi shown in Fig.~\ref{Figure::ResultsRpPb_prompt_nonprompt} (right). Also shown are results from a FONLL computation employing EPPS16 nPDFs~\cite{EPPS16_2017}, whose uncertainties strongly increase with decreasing \pt. 

\section{Summary}\label{Section::Summary}

This article presents for the first time the \pt-differential production cross sections and nuclear modification factors \RpPb of inclusive, prompt and non-prompt \jpsi at midrapidity ($ -1.37 < y_{\rm cms} < 0.43$) in \pPb collisions at $\sqrt{s_{\rm NN}} = 8.16$ TeV with the ALICE detector at the LHC. The measurements were made possible by the usage of online single-electron triggers provided by the Transition Radiation Detector.  Within the experimental and theoretical model uncertainties, the \pt-differential cross sections are well described by the calculations assuming only nuclear modified PDFs, with no final-state effects. The consistency of the measured \RpPb value of prompt \jpsi with unity shows that in the studied kinematic range cold nuclear matter effects are modest, even smaller than predicted by most theoretical calculations.

%%%%%%%%%%%%%%%%%%%%%%%%%%%%%%%%
% end main text 
%%%%%%%%%%%%%%%%%%%%%%%%%%%%%%%%

%%%%% acknowledgements - handled by EB chairs 
\newenvironment{acknowledgement}{\relax}{\relax}
\begin{acknowledgement}
\section*{Acknowledgements}
% add specific acknowledgements here 
% ...but please don't remove the line below: funding agencies
% will be acknowledged with a custom tex file handled by EB chairs after Collab Round 2
% Version: 2022-11-08

The ALICE Collaboration would like to thank all its engineers and technicians for their invaluable contributions to the construction of the experiment and the CERN accelerator teams for the outstanding performance of the LHC complex.
The ALICE Collaboration gratefully acknowledges the resources and support provided by all Grid centres and the Worldwide LHC Computing Grid (WLCG) collaboration.
The ALICE Collaboration acknowledges the following funding agencies for their support in building and running the ALICE detector:
A. I. Alikhanyan National Science Laboratory (Yerevan Physics Institute) Foundation (ANSL), State Committee of Science and World Federation of Scientists (WFS), Armenia;
Austrian Academy of Sciences, Austrian Science Fund (FWF): [M 2467-N36] and Nationalstiftung f\"{u}r Forschung, Technologie und Entwicklung, Austria;
Ministry of Communications and High Technologies, National Nuclear Research Center, Azerbaijan;
Conselho Nacional de Desenvolvimento Cient\'{\i}fico e Tecnol\'{o}gico (CNPq), Financiadora de Estudos e Projetos (Finep), Funda\c{c}\~{a}o de Amparo \`{a} Pesquisa do Estado de S\~{a}o Paulo (FAPESP) and Universidade Federal do Rio Grande do Sul (UFRGS), Brazil;
Bulgarian Ministry of Education and Science, within the National Roadmap for Research Infrastructures 2020-2027 (object CERN), Bulgaria;
Ministry of Education of China (MOEC) , Ministry of Science \& Technology of China (MSTC) and National Natural Science Foundation of China (NSFC), China;
Ministry of Science and Education and Croatian Science Foundation, Croatia;
Centro de Aplicaciones Tecnol\'{o}gicas y Desarrollo Nuclear (CEADEN), Cubaenerg\'{\i}a, Cuba;
Ministry of Education, Youth and Sports of the Czech Republic, Czech Republic;
The Danish Council for Independent Research | Natural Sciences, the VILLUM FONDEN and Danish National Research Foundation (DNRF), Denmark;
Helsinki Institute of Physics (HIP), Finland;
Commissariat \`{a} l'Energie Atomique (CEA) and Institut National de Physique Nucl\'{e}aire et de Physique des Particules (IN2P3) and Centre National de la Recherche Scientifique (CNRS), France;
Bundesministerium f\"{u}r Bildung und Forschung (BMBF) and GSI Helmholtzzentrum f\"{u}r Schwerionenforschung GmbH, Germany;
General Secretariat for Research and Technology, Ministry of Education, Research and Religions, Greece;
National Research, Development and Innovation Office, Hungary;
Department of Atomic Energy Government of India (DAE), Department of Science and Technology, Government of India (DST), University Grants Commission, Government of India (UGC) and Council of Scientific and Industrial Research (CSIR), India;
National Research and Innovation Agency - BRIN, Indonesia;
Istituto Nazionale di Fisica Nucleare (INFN), Italy;
Japanese Ministry of Education, Culture, Sports, Science and Technology (MEXT) and Japan Society for the Promotion of Science (JSPS) KAKENHI, Japan;
Consejo Nacional de Ciencia (CONACYT) y Tecnolog\'{i}a, through Fondo de Cooperaci\'{o}n Internacional en Ciencia y Tecnolog\'{i}a (FONCICYT) and Direcci\'{o}n General de Asuntos del Personal Academico (DGAPA), Mexico;
Nederlandse Organisatie voor Wetenschappelijk Onderzoek (NWO), Netherlands;
The Research Council of Norway, Norway;
Commission on Science and Technology for Sustainable Development in the South (COMSATS), Pakistan;
Pontificia Universidad Cat\'{o}lica del Per\'{u}, Peru;
Ministry of Education and Science, National Science Centre and WUT ID-UB, Poland;
Korea Institute of Science and Technology Information and National Research Foundation of Korea (NRF), Republic of Korea;
Ministry of Education and Scientific Research, Institute of Atomic Physics, Ministry of Research and Innovation and Institute of Atomic Physics and University Politehnica of Bucharest, Romania;
Ministry of Education, Science, Research and Sport of the Slovak Republic, Slovakia;
National Research Foundation of South Africa, South Africa;
Swedish Research Council (VR) and Knut \& Alice Wallenberg Foundation (KAW), Sweden;
European Organization for Nuclear Research, Switzerland;
Suranaree University of Technology (SUT), National Science and Technology Development Agency (NSTDA), Thailand Science Research and Innovation (TSRI) and National Science, Research and Innovation Fund (NSRF), Thailand;
Turkish Energy, Nuclear and Mineral Research Agency (TENMAK), Turkey;
National Academy of  Sciences of Ukraine, Ukraine;
Science and Technology Facilities Council (STFC), United Kingdom;
National Science Foundation of the United States of America (NSF) and United States Department of Energy, Office of Nuclear Physics (DOE NP), United States of America.
In addition, individual groups or members have received support from:
Marie Sk\l{}odowska Curie, European Research Council, Strong 2020 - Horizon 2020 (grant nos. 950692, 824093, 896850), European Union;
Academy of Finland (Center of Excellence in Quark Matter) (grant nos. 346327, 346328), Finland;
Programa de Apoyos para la Superaci\'{o}n del Personal Acad\'{e}mico, UNAM, Mexico.

\end{acknowledgement}

%%%%%%%% Bibliography 
\bibliographystyle{utphys}   % Remember we use title in the biblio
\bibliography{bibliography}
%\input {bibliography.tex}  

%%%%%%%%%%%%%%%%%%%%%%%%%%%%%%%%
% Appendices: yours (if any) + authorlist
%%%%%%%%%%%%%%%%%%%%%%%%%%%%%%%%
\newpage
\appendix

%
%\input{} % put your appendices here (if any)
%

%%%%% Authorlist - please do not touch: handled by EB chairs 
\section{The ALICE Collaboration}
\label{app:collab}
% ALICE Collaboration author list for 2022-11-08
\begin{flushleft} 
\small

S.~Acharya\,\orcidlink{0000-0002-9213-5329}\,$^{\rm 125}$, 
D.~Adamov\'{a}\,\orcidlink{0000-0002-0504-7428}\,$^{\rm 86}$, 
A.~Adler$^{\rm 69}$, 
G.~Aglieri Rinella\,\orcidlink{0000-0002-9611-3696}\,$^{\rm 32}$, 
M.~Agnello\,\orcidlink{0000-0002-0760-5075}\,$^{\rm 29}$, 
N.~Agrawal\,\orcidlink{0000-0003-0348-9836}\,$^{\rm 50}$, 
Z.~Ahammed\,\orcidlink{0000-0001-5241-7412}\,$^{\rm 132}$, 
S.~Ahmad\,\orcidlink{0000-0003-0497-5705}\,$^{\rm 15}$, 
S.U.~Ahn\,\orcidlink{0000-0001-8847-489X}\,$^{\rm 70}$, 
I.~Ahuja\,\orcidlink{0000-0002-4417-1392}\,$^{\rm 37}$
% ...

\section*{Affiliation Notes}

$^{\rm I}$ Deceased\\
$^{\rm II}$ Also at: Max-Planck-Institut f\"{u}r Physik, Munich, Germany\\
$^{\rm III}$ Also at: Italian National Agency for New Technologies, Energy and Sustainable Economic Development (ENEA), Bologna, Italy\\
$^{\rm IV}$ Also at: Dipartimento DET del Politecnico di Torino, Turin, Italy\\
$^{\rm V}$ Also at: Department of Applied Physics, Aligarh Muslim University, Aligarh, India\\
$^{\rm VI}$ Also at: Institute of Theoretical Physics, University of Wroclaw, Poland\\
$^{\rm VII}$ Also at: An institution covered by a cooperation agreement with CERN\\

\section*{Collaboration Institutes}

$^{1}$ A.I. Alikhanyan National Science Laboratory (Yerevan Physics Institute) Foundation, Yerevan, Armenia\\
$^{2}$ AGH University of Science and Technology, Cracow, Poland\\
$^{3}$ Bogolyubov Institute for Theoretical Physics, National Academy of Sciences of Ukraine, Kiev, Ukraine\\
$^{4}$ Bose Institute, Department of Physics  and Centre for Astroparticle Physics and Space Science (CAPSS), Kolkata, India\\
$^{5}$ California Polytechnic State University, San Luis Obispo, California, United States\\
$^{6}$ Central China Normal University, Wuhan, China\\
$^{7}$ Centro de Aplicaciones Tecnol\'{o}gicas y Desarrollo Nuclear (CEADEN), Havana, Cuba\\
$^{8}$ Centro de Investigaci\'{o}n y de Estudios Avanzados (CINVESTAV), Mexico City and M\'{e}rida, Mexico\\
$^{9}$ Chicago State University, Chicago, Illinois, United States\\
$^{10}$ China Institute of Atomic Energy, Beijing, China\\
$^{11}$ Chungbuk National University, Cheongju, Republic of Korea\\
$^{12}$ Comenius University Bratislava, Faculty of Mathematics, Physics and Informatics, Bratislava, Slovak Republic\\
$^{13}$ COMSATS University Islamabad, Islamabad, Pakistan\\
$^{14}$ Creighton University, Omaha, Nebraska, United States\\
$^{15}$ Department of Physics, Aligarh Muslim University, Aligarh, India\\
$^{16}$ Department of Physics, Pusan National University, Pusan, Republic of Korea\\
$^{17}$ Department of Physics, Sejong University, Seoul, Republic of Korea\\
$^{18}$ Department of Physics, University of California, Berkeley, California, United States\\
$^{19}$ Department of Physics, University of Oslo, Oslo, Norway\\
$^{20}$ Department of Physics and Technology, University of Bergen, Bergen, Norway\\
$^{21}$ Dipartimento di Fisica, Universit\`{a} di Pavia, Pavia, Italy\\
$^{22}$ Dipartimento di Fisica dell'Universit\`{a} and Sezione INFN, Cagliari, Italy\\
$^{23}$ Dipartimento di Fisica dell'Universit\`{a} and Sezione INFN, Trieste, Italy\\
$^{24}$ Dipartimento di Fisica dell'Universit\`{a} and Sezione INFN, Turin, Italy\\
$^{25}$ Dipartimento di Fisica e Astronomia dell'Universit\`{a} and Sezione INFN, Bologna, Italy\\
$^{26}$ Dipartimento di Fisica e Astronomia dell'Universit\`{a} and Sezione INFN, Catania, Italy\\
$^{27}$ Dipartimento di Fisica e Astronomia dell'Universit\`{a} and Sezione INFN, Padova, Italy\\
$^{28}$ Dipartimento di Fisica `E.R.~Caianiello' dell'Universit\`{a} and Gruppo Collegato INFN, Salerno, Italy\\
$^{29}$ Dipartimento DISAT del Politecnico and Sezione INFN, Turin, Italy\\
$^{30}$ Dipartimento di Scienze MIFT, Universit\`{a} di Messina, Messina, Italy\\
$^{31}$ Dipartimento Interateneo di Fisica `M.~Merlin' and Sezione INFN, Bari, Italy\\
$^{32}$ European Organization for Nuclear Research (CERN), Geneva, Switzerland\\
$^{33}$ Faculty of Electrical Engineering, Mechanical Engineering and Naval Architecture, University of Split, Split, Croatia\\
$^{34}$ Faculty of Engineering and Science, Western Norway University of Applied Sciences, Bergen, Norway\\
$^{35}$ Faculty of Nuclear Sciences and Physical Engineering, Czech Technical University in Prague, Prague, Czech Republic\\
$^{36}$ Faculty of Physics, Sofia University, Sofia, Bulgaria\\
$^{37}$ Faculty of Science, P.J.~\v{S}af\'{a}rik University, Ko\v{s}ice, Slovak Republic\\
$^{38}$ Frankfurt Institute for Advanced Studies, Johann Wolfgang Goethe-Universit\"{a}t Frankfurt, Frankfurt, Germany\\
$^{39}$ Fudan University, Shanghai, China\\
$^{40}$ Gangneung-Wonju National University, Gangneung, Republic of Korea\\
$^{41}$ Gauhati University, Department of Physics, Guwahati, India\\
$^{42}$ Helmholtz-Institut f\"{u}r Strahlen- und Kernphysik, Rheinische Friedrich-Wilhelms-Universit\"{a}t Bonn, Bonn, Germany\\
$^{43}$ Helsinki Institute of Physics (HIP), Helsinki, Finland\\
$^{44}$ High Energy Physics Group,  Universidad Aut\'{o}noma de Puebla, Puebla, Mexico\\
$^{45}$ Horia Hulubei National Institute of Physics and Nuclear Engineering, Bucharest, Romania\\
$^{46}$ Indian Institute of Technology Bombay (IIT), Mumbai, India\\
$^{47}$ Indian Institute of Technology Indore, Indore, India\\
$^{48}$ INFN, Laboratori Nazionali di Frascati, Frascati, Italy\\
$^{49}$ INFN, Sezione di Bari, Bari, Italy\\
$^{50}$ INFN, Sezione di Bologna, Bologna, Italy\\
$^{51}$ INFN, Sezione di Cagliari, Cagliari, Italy\\
$^{52}$ INFN, Sezione di Catania, Catania, Italy\\
$^{53}$ INFN, Sezione di Padova, Padova, Italy\\
$^{54}$ INFN, Sezione di Pavia, Pavia, Italy\\
$^{55}$ INFN, Sezione di Torino, Turin, Italy\\
$^{56}$ INFN, Sezione di Trieste, Trieste, Italy\\
$^{57}$ Inha University, Incheon, Republic of Korea\\
$^{58}$ Institute for Gravitational and Subatomic Physics (GRASP), Utrecht University/Nikhef, Utrecht, Netherlands\\
$^{59}$ Institute of Experimental Physics, Slovak Academy of Sciences, Ko\v{s}ice, Slovak Republic\\
$^{60}$ Institute of Physics, Homi Bhabha National Institute, Bhubaneswar, India\\
$^{61}$ Institute of Physics of the Czech Academy of Sciences, Prague, Czech Republic\\
$^{62}$ Institute of Space Science (ISS), Bucharest, Romania\\
$^{63}$ Institut f\"{u}r Kernphysik, Johann Wolfgang Goethe-Universit\"{a}t Frankfurt, Frankfurt, Germany\\
$^{64}$ Instituto de Ciencias Nucleares, Universidad Nacional Aut\'{o}noma de M\'{e}xico, Mexico City, Mexico\\
$^{65}$ Instituto de F\'{i}sica, Universidade Federal do Rio Grande do Sul (UFRGS), Porto Alegre, Brazil\\
$^{66}$ Instituto de F\'{\i}sica, Universidad Nacional Aut\'{o}noma de M\'{e}xico, Mexico City, Mexico\\
$^{67}$ iThemba LABS, National Research Foundation, Somerset West, South Africa\\
$^{68}$ Jeonbuk National University, Jeonju, Republic of Korea\\
$^{69}$ Johann-Wolfgang-Goethe Universit\"{a}t Frankfurt Institut f\"{u}r Informatik, Fachbereich Informatik und Mathematik, Frankfurt, Germany\\
$^{70}$ Korea Institute of Science and Technology Information, Daejeon, Republic of Korea\\
$^{71}$ KTO Karatay University, Konya, Turkey\\
$^{72}$ Laboratoire de Physique des 2 Infinis, Ir\`{e}ne Joliot-Curie, Orsay, France\\
$^{73}$ Laboratoire de Physique Subatomique et de Cosmologie, Universit\'{e} Grenoble-Alpes, CNRS-IN2P3, Grenoble, France\\
$^{74}$ Lawrence Berkeley National Laboratory, Berkeley, California, United States\\
$^{75}$ Lund University Department of Physics, Division of Particle Physics, Lund, Sweden\\
$^{76}$ Nagasaki Institute of Applied Science, Nagasaki, Japan\\
$^{77}$ Nara Women{'}s University (NWU), Nara, Japan\\
$^{78}$ National and Kapodistrian University of Athens, School of Science, Department of Physics , Athens, Greece\\
$^{79}$ National Centre for Nuclear Research, Warsaw, Poland\\
$^{80}$ National Institute of Science Education and Research, Homi Bhabha National Institute, Jatni, India\\
$^{81}$ National Nuclear Research Center, Baku, Azerbaijan\\
$^{82}$ National Research and Innovation Agency - BRIN, Jakarta, Indonesia\\
$^{83}$ Niels Bohr Institute, University of Copenhagen, Copenhagen, Denmark\\
$^{84}$ Nikhef, National institute for subatomic physics, Amsterdam, Netherlands\\
$^{85}$ Nuclear Physics Group, STFC Daresbury Laboratory, Daresbury, United Kingdom\\
$^{86}$ Nuclear Physics Institute of the Czech Academy of Sciences, Husinec-\v{R}e\v{z}, Czech Republic\\
$^{87}$ Oak Ridge National Laboratory, Oak Ridge, Tennessee, United States\\
$^{88}$ Ohio State University, Columbus, Ohio, United States\\
$^{89}$ Physics department, Faculty of science, University of Zagreb, Zagreb, Croatia\\
$^{90}$ Physics Department, Panjab University, Chandigarh, India\\
$^{91}$ Physics Department, University of Jammu, Jammu, India\\
$^{92}$ Physics Program and International Institute for Sustainability with Knotted Chiral Meta Matter (SKCM2), Hiroshima University, Hiroshima, Japan\\
$^{93}$ Physikalisches Institut, Eberhard-Karls-Universit\"{a}t T\"{u}bingen, T\"{u}bingen, Germany\\
$^{94}$ Physikalisches Institut, Ruprecht-Karls-Universit\"{a}t Heidelberg, Heidelberg, Germany\\
$^{95}$ Physik Department, Technische Universit\"{a}t M\"{u}nchen, Munich, Germany\\
$^{96}$ Politecnico di Bari and Sezione INFN, Bari, Italy\\
$^{97}$ Research Division and ExtreMe Matter Institute EMMI, GSI Helmholtzzentrum f\"ur Schwerionenforschung GmbH, Darmstadt, Germany\\
$^{98}$ Saga University, Saga, Japan\\
$^{99}$ Saha Institute of Nuclear Physics, Homi Bhabha National Institute, Kolkata, India\\
$^{100}$ School of Physics and Astronomy, University of Birmingham, Birmingham, United Kingdom\\
$^{101}$ Secci\'{o}n F\'{\i}sica, Departamento de Ciencias, Pontificia Universidad Cat\'{o}lica del Per\'{u}, Lima, Peru\\
$^{102}$ Stefan Meyer Institut f\"{u}r Subatomare Physik (SMI), Vienna, Austria\\
$^{103}$ SUBATECH, IMT Atlantique, Nantes Universit\'{e}, CNRS-IN2P3, Nantes, France\\
$^{104}$ Sungkyunkwan University, Suwon City, Republic of Korea\\
$^{105}$ Suranaree University of Technology, Nakhon Ratchasima, Thailand\\
$^{106}$ Technical University of Ko\v{s}ice, Ko\v{s}ice, Slovak Republic\\
$^{107}$ The Henryk Niewodniczanski Institute of Nuclear Physics, Polish Academy of Sciences, Cracow, Poland\\
$^{108}$ The University of Texas at Austin, Austin, Texas, United States\\
$^{109}$ Universidad Aut\'{o}noma de Sinaloa, Culiac\'{a}n, Mexico\\
$^{110}$ Universidade de S\~{a}o Paulo (USP), S\~{a}o Paulo, Brazil\\
$^{111}$ Universidade Estadual de Campinas (UNICAMP), Campinas, Brazil\\
$^{112}$ Universidade Federal do ABC, Santo Andre, Brazil\\
$^{113}$ University of Cape Town, Cape Town, South Africa\\
$^{114}$ University of Houston, Houston, Texas, United States\\
$^{115}$ University of Jyv\"{a}skyl\"{a}, Jyv\"{a}skyl\"{a}, Finland\\
$^{116}$ University of Kansas, Lawrence, Kansas, United States\\
$^{117}$ University of Liverpool, Liverpool, United Kingdom\\
$^{118}$ University of Science and Technology of China, Hefei, China\\
$^{119}$ University of South-Eastern Norway, Kongsberg, Norway\\
$^{120}$ University of Tennessee, Knoxville, Tennessee, United States\\
$^{121}$ University of the Witwatersrand, Johannesburg, South Africa\\
$^{122}$ University of Tokyo, Tokyo, Japan\\
$^{123}$ University of Tsukuba, Tsukuba, Japan\\
$^{124}$ University Politehnica of Bucharest, Bucharest, Romania\\
$^{125}$ Universit\'{e} Clermont Auvergne, CNRS/IN2P3, LPC, Clermont-Ferrand, France\\
$^{126}$ Universit\'{e} de Lyon, CNRS/IN2P3, Institut de Physique des 2 Infinis de Lyon, Lyon, France\\
$^{127}$ Universit\'{e} de Strasbourg, CNRS, IPHC UMR 7178, F-67000 Strasbourg, France, Strasbourg, France\\
$^{128}$ Universit\'{e} Paris-Saclay Centre d'Etudes de Saclay (CEA), IRFU, D\'{e}partment de Physique Nucl\'{e}aire (DPhN), Saclay, France\\
$^{129}$ Universit\`{a} degli Studi di Foggia, Foggia, Italy\\
$^{130}$ Universit\`{a} del Piemonte Orientale, Vercelli, Italy\\
$^{131}$ Universit\`{a} di Brescia, Brescia, Italy\\
$^{132}$ Variable Energy Cyclotron Centre, Homi Bhabha National Institute, Kolkata, India\\
$^{133}$ Warsaw University of Technology, Warsaw, Poland\\
$^{134}$ Wayne State University, Detroit, Michigan, United States\\
$^{135}$ Westf\"{a}lische Wilhelms-Universit\"{a}t M\"{u}nster, Institut f\"{u}r Kernphysik, M\"{u}nster, Germany\\
$^{136}$ Wigner Research Centre for Physics, Budapest, Hungary\\
$^{137}$ Yale University, New Haven, Connecticut, United States\\
$^{138}$ Yonsei University, Seoul, Republic of Korea\\
$^{139}$  Zentrum  f\"{u}r Technologie und Transfer (ZTT), Worms, Germany\\
$^{140}$ Affiliated with an institute covered by a cooperation agreement with CERN\\
$^{141}$ Affiliated with an international laboratory covered by a cooperation agreement with CERN.\\

\end{flushleft} 
  
\end{document}